\documentclass[]{aa} 
\usepackage{txfonts}
\usepackage{natbib}
\usepackage{graphicx, xspace}
\usepackage{placeins}
\usepackage{array}
\usepackage[switch]{lineno}
\usepackage{hyperref}
\usepackage{color}
\usepackage{gensymb}
\usepackage{amsmath}

\newcommand{\Lagr}{\mathcal{L}}
\newcommand{\Magr}{\mathcal{M}}
\newcommand{\Dagr}{\mathcal{D}}

\makeatletter

\newcommand{\Rmnum}[1]{\expandafter\@slowromancap\romannumeral #1@}
\makeatother

\def\d{\,d$^{-1}$\xspace}

\def\ZetOri{$\zeta$\,Ori\xspace}  

\begin{document}
\title{Studying the photometric and spectroscopic variability of the magnetic hot supergiant $\zeta$\,Orionis\,Aa 
\thanks{Based on data collected by the BRITE Constellation satellite mission, designed, built, launched, operated and supported by the Austrian Research Promotion Agency (FFG), the University of Vienna, the Technical University of Graz, the Canadian Space Agency (CSA), the University of Toronto Institute for Aerospace Studies (UTIAS), the Foundation for Polish Science \& Technology (FNiTP MNiSW), and National Science Centre (NCN).}$^{, }$
\thanks{Based on CHIRON spectra collected under CNTAC proposal CN2015A-122.}}

\authorrunning{B.\,Buysschaert}
\titlerunning{Study of $\zeta$\,Orionis\,Aa}

\author{B.\,Buysschaert\inst{1,2}, 
C.\,Neiner\inst{1},
N.\,D.\,Richardson\inst{3},
T.\,Ramiaramanantsoa\inst{4},
A.\,David-Uraz\inst{5},
H.\,Pablo\inst{4},
M.\,E.\,Oksala\inst{1, 6},
A.\,F.\,J.\,Moffat\inst{4},
R.\,E.\,Mennickent\inst{7},
S.\,Legeza\inst{3},
C.\,Aerts\inst{2,8},
R.\,Kuschnig\inst{9, 10},
G.N.\,Whittaker\inst{11},
A.\,Popowicz\inst{12},
G.\,Handler\inst{13},
G.\,A.\,Wade\inst{14},
W.\,W.\,Weiss\inst{10}\\
}
\mail{bram.buysschaert@obspm.fr} 

\institute{
LESIA, Observatoire de Paris, PSL Research University, CNRS, Sorbonne Universit\'es, UPMC Univ. Paris 06, Univ. Paris Diderot, Sorbonne Paris Cit\'e, 5 place Jules Janssen, F-92195 Meudon, France 
\and Instituut voor Sterrenkunde, KU Leuven, Celestijnenlaan 200D, 3001 Leuven, Belgium 
\and Ritter Observatory, Department of Physics and Astronomy, The University of Toledo, Toledo, OH 43606-3390, USA 
\and D\'epartement de physique and Centre de Recherche en Astrophysique du Qu\'ebec (CRAQ), Universit\'e de Montr\'eal, C.P. 6128,
Succ. Centre-Ville, Montr\'eal, Qu\'ebec, H3C 3J7, Canada 
\and Department of Physics \& Space Sciences, Florida Institute of Technology, Melbourne, FL, 32901, USA 
\and Department of Physics, California Lutheran University, 60 West Olsen Road 3700, Thousand Oaks, CA, 91360 
\and Departamento de Astronom{\'{i}}a, Universidad de Concepci\'on, Casilla 160-C, Concepci\'on, Chile 
\and Dept. of Astrophysics, IMAPP, Radboud University Nijmegen, 6500 GL, Nijmegen, The Netherlands 
\and Graz University of Technology, Institute of Communication Networks and Satellite Communications, Inffeldgasse 12, 8010 Graz, Austria 
\and Institut f\"ur Astrophysik, Universit\"at Wien, T\"urkenschanzstrasse 17, 1180 Wien, Austria 
\and Department of Astronomy \& Astrophysics, University of Toronto, 50 St. George St, Toronto, Ontario, M5S 3H4, Canada 
\and Silesian University of Technology, Institute of Automatic Control, Gliwice, Akademicka 16, Poland 
\and Centrum Astronomiczne im. M. Kopernika, Polska Akademia Nauk, Bartycka 18, 00-716 Warszawa, Poland 
\and Department of Physics, Royal Military College of Canada, PO Box 17000, Station Forces, Kingston, Ontario, K7K 7B4, Canada 
}
\abstract{
Massive stars play a significant role in the chemical and dynamical evolution of galaxies. However, much of their variability, particularly during their evolved supergiant stage, is poorly understood.  To understand the variability of evolved massive stars in more detail, we present a study of the O9.2Ib supergiant \ZetOri\,Aa, the only currently confirmed supergiant to host a magnetic field.  We have obtained two-color space-based BRIght Target Explorer photometry (BRITE) for \ZetOri\,Aa during two observing campaigns, as well as simultaneous ground-based, high-resolution optical CHIRON spectroscopy. We perform a detailed frequency analysis to detect and characterize the star's periodic variability.  We detect two significant, independent frequencies, their higher harmonics, and combination frequencies: the stellar rotation period $P_{\mathrm{rot}}\,=\,6.82\pm0.18$\,d, most likely related to the presence of the stable magnetic poles, and a variation with a period of 10.0$\pm$0.3\,d attributed to circumstellar environment, also detected in the H$\alpha$ and several He\,I lines, yet absent in the purely photospheric lines. We confirm the variability with $P_{\mathrm{rot}}$/4, likely caused by surface inhomogeneities, being the possible photospheric drivers of the discrete absorption components. No stellar pulsations were detected in the data. The level of circumstellar activity clearly differs between the two BRITE observing campaigns. We demonstrate that \ZetOri\,Aa is a highly variable star with both periodic and non-periodic variations, as well as episodic events. The rotation period we determined agrees well with the spectropolarimetric value from the literature. The changing activity level observed with BRITE could explain why the rotational modulation of the magnetic measurements was not clearly detected at all epochs.}


\maketitle

\section{Introduction}
\label{sec:Introduction}
Massive stars ($M_{init} > 8M_{\odot}$) are the chemical factories of the Universe, as they contribute significantly to its chemical enrichment \citep[e.g.,][]{2009pfer.book.....M}. Thanks to their high mass, they reach the necessary conditions to undergo all stages of nuclear burning. Moreover, their lifetimes are much shorter than those of their less massive counterparts, characterized by various epochs of considerable mass loss, after which the stars come to an abrupt and explosive ending. Massive stars \citep[below 50 $M_\odot$,][]{2017A+A...598A..56M} evolve to the supergiant stage when core hydrogen burning has ended. Such stars have variability of various origins, often related to or caused by different physical effects, which we discuss below.

Massive supergiants undergo strong mass loss through line driven winds \citep{2000ARA+A..38..613K}. The properties of the wind and the associated mass loss were studied for many massive supergiant stars by investigating the P\,Cygni profiles of the \ion{Si}{IV} and \ion{N}{V} lines observed by the International Ultraviolet Explorer \citep[IUE; e.g.,][]{2002A+A...388..587P}. These observations indicated the presence of time-variable absorption structures, known as discrete absorption components (DACs), inferred to exist for most massive OB stars \citep[e.g.,][]{1989ApJS...69..527H, 1996A+AS..116..257K}. Such structures start as broad low-velocity components close to the stellar surface and evolve to narrow high-velocity components. The recurrence timescale and the acceleration timescale of these structures both correlate with the rotation velocity of the star, yet the strength between consecutive rotation cycles varies, as does their coherence on longer timescales.  Such variations are defined to be cyclic rather than periodic, since the amplitude of the variations differs on a timescale of several stellar rotations.

\citet{1984ApJ...283..303M, 1986A+A...165..157M} proposed that corotating interaction regions (CIRs), i.e., spiral-shaped wind structures in corotation with the star, lay at the origin of DACs. Theoretical work by \citet{1996ApJ...462..469C} showed that localized perturbations at the stellar surface, such as bright spots, produce CIRs in the star's circumstellar environment, and that slowly propagating discontinuous velocity plateaus in the wind located ahead of the compressed CIR arms cause the DACs, instead of the CIRs themselves. Recent high-precision photometric monitoring of the mid-O-type giant $\xi$~Per by the Microvariability and Oscillations of STars (MOST) microsatellite shows evidence of signatures of corotating bright spots at the surface of the star, with a potential link to its DAC behavior \citep{2014MNRAS.441..910R}. Still, the origin of the surface inhomogeneities remains uncertain; magnetic fields \citep[e.g.,][]{1994IAUS..162..517H} and non-radial pulsations \citep[e.g.,][]{1986PASP...98...37W, 1986PASP...98...52C} have been proposed as scenarios.

Large-scale magnetic fields continue to be the preferred pathway to produce local surface anisotropies, since they can cause chemical stratification or lower the optical depth, due to magnetic pressure at the magnetic poles, commonly leading to surface brightness spots. Moreover, they also naturally explain differences in the wind velocity, aiding the formation of CIRs. However, a recent study by \citet{2014MNRAS.444..429D} detected no large-scale magnetic fields for a sample of mainly OB supergiants having a well-studied DAC behavior.  Their derived upper limits on non-detected polar fields was typically larger than 50\,G.  These limits are larger than the typical field strengths expected under the assumption of magnetic flux conservation for the increased stellar radius during the supergiant phase \citep{2006A+A...450..777B, 2007A+A...470..685L, 2008A+A...481..465L, 2016A+A...592A..84F}, which only reaches up to a few tens of gauss.  Additionally, the determined upper limits on the strength of the magnetic dipole do not reach the critical strength needed to significantly perturb the wind.

A different approach has been proposed by \citet{2011A+A...534A.140C}, in which small-scale magnetic fields caused by a near-surface convective layer produce hot, bright spots.  Theoretical calculations indicate that the strength of such small-scale magnetic fields ranges between 5\,G and 50\,G and would lead to a very complex circular polarization signature.  While these magnetic fields could in principle generate the surface spots needed to explain the almost ubiquitous presence of DACs, none have so far been detected for massive OB stars during large surveys \citep[e.g., the MiMeS survey;][]{2016MNRAS.456....2W} and ultra-deep searches \citep[e.g.,][which reach a sub-gauss precision]{2014A+A...562A..59N, 2014MNRAS.444.1993W}.

Some hot supergiants also exhibit stellar pulsations. These prove to be crucial ingredients to unraveling the internal properties of such objects, since pulsations are directly related to internal stellar conditions. Different types of pulsations are theoretically expected, yet only scarcely detected in massive evolved stars \citep[see][for a monograph on asteroseismology]{2010aste.book.....A}. 

Slowly pulsating B-star (SPB) oscillations are expected to occur in OB supergiants. Such pulsations are driven by the $\kappa$-mechanism acting on the iron opacity bump at $\mathrm{T}\,\sim\,2\times\,10^5$\,K. However, an intermediate convective layer just above the hydrogen burning shell is needed to prevent the gravity modes from propagating into the core \citep{2009A+A...498..273G, 2009MNRAS.396.1833G}. These SPB supergiants have low-degree, high-order gravity pulsation modes, with a period of several days \citep[e.g.,][]{2006ApJ...650.1111S, 2007A+A...463.1093L}.

Oscillatory convection (non-adiabatic $\rm g^{-}$) modes have been theoretically predicted to occur at the stellar surface of massive supergiants, with periods ranging from about $1.5$\,days to hundreds of days \citep[][]{2011MNRAS.412.1814S}. Low-degree oscillatory convection modes are driven at the convective envelope associated with the iron-group opacity peak, but only become overstable if non-adiabatic effects are included \citep{1981PASJ...33..427S}. These modes are believed to cause some of the variability seen in massive supergiants, e.g., $\zeta$\,Pup \citep[O4I(n)fp;][]{2014MNRAS.445.2878H}.  On the other hand, the discovered single 1.78\,d periodicity in this star has also been interpreted as caused by rotating photospheric bright spots instead of stellar pulsations (Ramiaramanantsoa et al., in prep.).

\citet{2012ApJ...749...74M} invoked gravity modes driven by fluctuations in the energy generation rate $\epsilon$ \citep[i.e., the $\epsilon$-mechanism;][]{1989nos..book.....U} to describe the longest variability periods determined for Rigel (B8Ia). This mechanism occurs when the heat exchange reaches a maximum. For massive supergiants, such as Rigel, these fluctuations take place in the hydrogen burning shell partly located in the radiative zone, just below the intermediate convection layer. 

Finally, strange-mode pulsations are theoretically expected for the most luminous massive supergiants \citep{1998MNRAS.294..622S, 2009CoAst.158..245S, 2009CoAst.158..252G, 2011IAUS..272..503G}. They are produced by the opacity mechanism on Fe and \ion{He}{II} ionization stages \citep{2011MNRAS.412.1814S}. However, none have yet been firmly detected.

Different types of stellar pulsations are thus theoretically expected, and on many occasions observed.  However, no firm link between these stellar pulsations and the origin of DACs has currently been noted.  As such, neither of the proposed photospheric driving mechanisms is clearly supported or disproved.  Therefore, it is possible that various processes produce surface brightness spots leading to DACs, for example non-radial pulsations (NRPs) or large-scale magnetic fields.

Variability on a short timescale, on the order of several minutes, has been noted for some supergiants \citep[e.g.,][]{2014MNRAS.440.1779H, 2017arXiv170100733K}.  The exact mechanism, however, remains uncertain, but it might be related to a stochastic process.

In this work, we study \ZetOri\,Aa, the only confirmed magnetic O-type supergiant to date. The star has previously shown complex behavior and variability, making it a very interesting object. We collected two-color space-based BRIght Target Explorer (BRITE) photometry as well as simultaneous ground-based high-resolution optical CHIRON spectroscopy, to study both the photometric and spectroscopic periodic variability of \ZetOri\,Aa by means of a detailed frequency analysis.  The frequency domain 0 -- 10\,\d is investigated since the signal of any periodic or cyclic magnetic, wind, circumstellar or stellar variability is expected in this frequency domain.

We first introduce \ZetOri\,Aa and the current understanding of the object in Sect.\,\ref{sec:zeta_ori}. Next, we describe the photometric observations and present a detailed frequency analysis in Sect.\,\ref{sec:BRITE}. Section\,\ref{sec:spectroscopy} contains the spectroscopic measurements and analysis. Our results are discussed in detail in Sect.\,\ref{sec:discussion}. Finally, we draw conclusions on the variability of \ZetOri\,Aa and make remarks for future studies in Sect.\,\ref{sec:conclusions}.

\begin{table}[t]
\caption{Stellar parameters for the three main components of \ZetOri, determined by \citet{2013A+A...554A..52H} unless noted differently. Stellar masses and radii were determined from the photometric distance.}
\centering
\tabcolsep=6pt
\begin{tabular}{p{1.5cm}ccc}
\hline
\hline
 & Aa & Ab & B\\
\hline
$\mathrm{m_V}$\,(mag) 			& 2.08					& 4.28			& 4.01		\\
SpT 								& O9.2Ib\,Nwk 	  $^{a}$	& B1IV			& B0III		\\
M\,($\mathrm{M_{\odot}}$) 		& $33 \pm 10$			& $14 \pm 3$		& --			\\
R\,($\mathrm{R_{\odot}}$) 		& $20 \pm 3.2$			& $7.3 \pm 1.0$	& --			\\
$v\sin i$\,($\mathrm{km/s}$) 	& 127		      $^{b}$	& $<$ 100 $^{c}$	& 350		\\
$\rm T_{eff}$\,(kK) 				& $29.5\pm1.0$    $^{d}$	& 29		  $^{c}$	& --			\\
$\log g$\,(dex) 					& $3.25 \pm 0.10$ $^{d}$	& 4.0	  $^{c}$	& --			\\
\hline
\end{tabular}
\label{tab:zetori_params}
\tablefoot{${a}$: \citet{2014ApJS..211...10S}; ${b}$: \citet{2014A+A...562A.135S}; ${c}$: \citet{2015A+A...582A.110B}; ${d}$: \citet{2008MNRAS.389...75B}.}
\end{table}

\section{$\zeta$\,Orionis}
\label{sec:zeta_ori}
\subsection{Multiplicity of $\zeta$\,Ori}
\label{sec:zetori_multiple}
\ZetOri (Alnitak, V\,=\,1.79) is a wide visual binary system consisting of \ZetOri\,A (HR\,1948) and \ZetOri\,B (HR\,1949), currently separated by $\sim2.4\,\arcsec$. The changes in common proper motion listed in the \textit{Washington Double Star Catalog} led to an orbital solution with a period longer than several hundred years \citep{2001AJ....122.3466M, 2008AJ....136..554T}. Since the components have such a large separation, it is assumed that they have not had any significant influence on each other's evolution.

More recently, \citet{2000ApJ...540L..91H} employed optical interferometry and showed that \ZetOri\,A is itself a binary system, composed of the massive supergiant \ZetOri\,Aa and an additional hot companion \ZetOri\,Ab 40\,mas away. The intensive spectroscopic and interferometric monitoring campaign permitted \citet{2013A+A...554A..52H} to deduce the orbital solution of the Aa+Ab system, together with some of the physical properties of both components. The system is moderately eccentric, $e\,=\,0.338\pm0.004$, with an orbital period of $2687.3\pm7.0$\,d ($\sim$\,7.4\,years). The long orbital period and eccentricity indicate a low probability of tidal interaction between the two components.  However, such interaction cannot be fully excluded, because of the large radius of \ZetOri\,Aa and the presence of a third component (\ZetOri\,B), which could hamper circularization of the system \citep{correia2012}.

Finally, a much fainter fourth component, \ZetOri\,C, is located about 57\,$\arcsec$ away from \ZetOri\,Aa. Because of the large brightness difference (>7\,mag) and its considerable separation, we ignore \ZetOri\,C for the remainder of this work. The physical parameters of the other three components are given in Table\,\ref{tab:zetori_params}.

\subsection{Magnetism}
\label{sec:zetori_magnetic}
\citet{2008MNRAS.389...75B} detected the possible presence of a weak magnetic field for the primary \ZetOri\,A. Using the Narval spectropolarimeter \citep{2003EAS.....9..105A}, the authors estimated the approximate rotation period of the star, and determined the geometrical configuration of the magnetic field. However, this study did not account for the component \ZetOri\,Ab, which at that time had not yet been explored \citep{2013A+A...554A..52H}. More recently, \citet{2015A+A...582A.110B} determined the rotation period and the magnetic configuration more precisely, by combining archival and new Narval spectropolarimetry, while accounting for the presence of \ZetOri\,Ab by performing spectral disentangling of the two components. This analysis led to a rotation period of $6.83 \pm 0.08$\,d for \ZetOri Aa (Blaz{\`e}re, private communication), and the confirmation of a dipolar magnetic field with a polar strength of about 140\,G. This rotation period is clearly visible in the 2007--8 magnetic measurements, but less clear in the 2011--12 data, because the latter was affected by additional unexplained variability. No magnetic field was detected for the companion \ZetOri\,Ab, with an upper limit for the polar field strength of 300\,G.

Furthermore, the presence of a weak dynamical magnetosphere was considered by \citet{2008MNRAS.389...75B} and \citet{2015A+A...582A.110B}, because of the locked $\rm{H}\alpha$ emission with the rotation period, an indication of magnetically confined circumstellar material. A compatible result was obtained by \citet{2004MNRAS.351..552M}, although they observed a lower harmonic of the rotation period. The large uncertainty on the mass-loss rate and the terminal wind speed makes a qualitative calculation of the Alfv\'en radius, and thus the precise confinement of the magnetosphere, difficult \citep[see][]{2015A+A...582A.110B}.

The studied X-ray observations of \ZetOri\,Aa are inconclusive, as different works favor \citep{2001ApJ...548L..45W} or argue against \citep{2014MNRAS.444.3729C, 2014ApJS..215...10N} the influence of the magnetic field on the X-ray variability of \ZetOri\,Aa. Since \ZetOri\,Aa hosts a weak magnetic field during a turbulent evolutionary stage, it is indeed likely that the weak magnetosphere only very weakly contributes to the X-ray variability, and that the emission instead originates from wind shocks. This stellar wind may have violent epochs; significant X-ray flux increases have been observed for \ZetOri\,Aa \citep{1994Sci...265.1689B}, although such variability is not always reported \citep[e.g.,][]{2014ApJS..215...10N}.

\subsection{Discrete absorption components}
\label{sec:zetori_dac}
Like most massive stars, \ZetOri\,A shows evidence of wind variability, manifested by DACs, in their UV resonance lines. \citet{1996A+AS..116..257K, 1999A+A...344..231K} studied this variability, using 29 UV spectra taken by IUE over a time span of 5.1\,days. The authors reported the presence of DACs for \ZetOri\,A, visible in both the \ion{Si}{IV}\,$\lambda\lambda1393.8,1402.8$ and \ion{N}{V}\,$\lambda\lambda1238.8,1242.8$ doublets with a recurrence timescale of $t_{\mathrm{rec}}\,=\,1.6\pm0.2$\,d. Unlike most stars with identified DACs, the recurrence timescale for \ZetOri\,A is roughly one-quarter of the inferred rotation period instead of the more typical one-half period. Considering the bright spot paradigm as the possible photospheric origin of the CIRs \citep[as in the canonical model of][]{1996ApJ...462..469C}, this means that four surface spots would be needed to create the accompanying CIRs in \ZetOri\,Aa, instead of two. Moreover, \citet{1999A+A...344..231K} state that the DACs are relatively weak for a supergiant, indicating there might indeed be something fundamentally different in the physical process for \ZetOri\,A. Finally, the authors observed a strong peak at a period of $\sim6$\,d in the Fourier analysis of their UV time series observations of \ZetOri\,A, but the interpretation of that periodicity was hampered by the fact that their observations only spanned $5.1$\,d. This $\sim6$\,d period can probably be associated with the rotation period ($6.83 \pm 0.08$\,d) determined from spectropolarimetry.

\section{BRITE photometry}
\label{sec:BRITE}
\begin{table*}[t]
\caption{Diagnostics related to the different BRITE observations of \ZetOri. For each setup, we provide information related to the cadence, the start time and length of the observations, the rms flux stability within an orbit passage before and after correction, the median duty cycle per satellite orbit, and the duty cycle of orbit passages where data were successfully taken. The duty cycles are determined for the corrected lightcurves. The two studied parts of the Orion\,II campaign, considered as different epochs, are marked as Orion\,IIa and Orion\,IIb (see Sect.\,\ref{sec:results_timeseries}).}
\centering
\tabcolsep=6pt
\begin{tabular}{p{1cm}p{1.3cm}lllllllll}
\hline
\hline
Satellite & Run & Setup & Stacked& $\mathrm{T_{start}}$ & Length & $\mathrm{RMS_{raw}}$ & $\mathrm{RMS_{corr}}$ & $\mathrm{D_{orb}}$ & $\mathrm{D_{sat}}$\\
&&Set&Observ.& [HJD-2450000] & [d] & [ppt] & [ppt] & [\%] & [\%]\\
\hline
BAb			&	Orion\,I		&	3	&	1	&	6628.43	&	74.01	&	1.76	&	0.86	&	11.7&	32.2\\
BAb			&	Orion\,I		&	4	&	1	&	6702.51	&	29.99	&	1.37	&	0.72	&	13.5&	66.5\\
UBr			&	Orion\,I		&	7	&	1	&	6603.61	&	129.28	&	1.00	&	0.64	&	14.2&	40.6\\\\
															
BAb			&	Orion\,II		&	2	&	1	&	6926.35	&	10.75	&	1.85	&	1.50	&	6.0&		50.6\\
BAb			&	Orion\,II		&	3	&	1	&	6937.37	&	7.55   	&	1.36	&	1.16	&	5.8&		40.6\\
BAb			&	Orion\,II		&	4	&	1	&	6945.39	&	21.64	&	3.81	&	1.08	&	6.7&		21.3\\
BTr			&	Orion\,II		&	1	&	1	&	6924.72	&	47.14	&	1.85	&	1.42	&	16.6&	70.8\\
BTr $^{a}$	&	Orion\,II		&	2	&	3	&	6972.17	&	2.88	    &	2.00	&	1.64	&	24.8&	89.9\\
BTr $^{b}$	&	Orion\,II		&	3	&	3	&	6987.61	&	7.99	    &	1.51	&	N/A	&	N/A&		N/A\\
BLb			&	Orion\,IIa	&	3	&	5	&	6998.52	&	45.04	&	1.45	&	1.19	&	12.1&	75.8\\
BLb			&	Orion\,IIb	&	6	&	1	&	7052.75	&	45.53	&	1.20	&	0.55&	17.0&	89.2\\
BHr $^{a}$	&	Orion\,II		&	2	&	3	&	6972.24	&	11.01	&	0.89	&	0.81	&	7.6&		59.4\\
BHr			&	Orion\,IIa	&	5	&	5	&	6998.56	&	49.06	&	1.05	&	0.85	&	12.5&	78.1\\
BHr	$^{c}$	&	Orion\,II		&	6	&	1	&	7049.56	&	7.30	    &	12.22&	N/A	&	N/A&		N/A\\
BHr			&	Orion\,IIb	&	7	&	1	&	7056.99	&	38.53	&	1.10	&	0.52	&	18.8&	67.7\\
\hline
\end{tabular}
\label{tab:BRITE_observations}
\tablefoot{The different BRITE nano-satellites are BAb (BRITE Austria blue), UBr (UniBRITE red), BTr (BRITE Toronto red), BLb (BRITE Lem blue), and BHr (BRITE Heweliusz red). ${a}$: After processing, there is a significant timelength reduction due to strong outliers. The timelength mentioned here is that of the corrected photometry. ${b}$: These observations were discarded because of i) the strong irregular temperature variability, and ii) the mismatch between the aperture and the CCD sub-raster. ${c}$: These observations were not used, due to the presence of strong charge-transfer-inefficiency (CTI) effects. N/A: Not applicable, since the value was not calculated.}
\end{table*}

\begin{figure*}[t]
		\centering
			\includegraphics[width=\textwidth, height = 0.33\textheight]{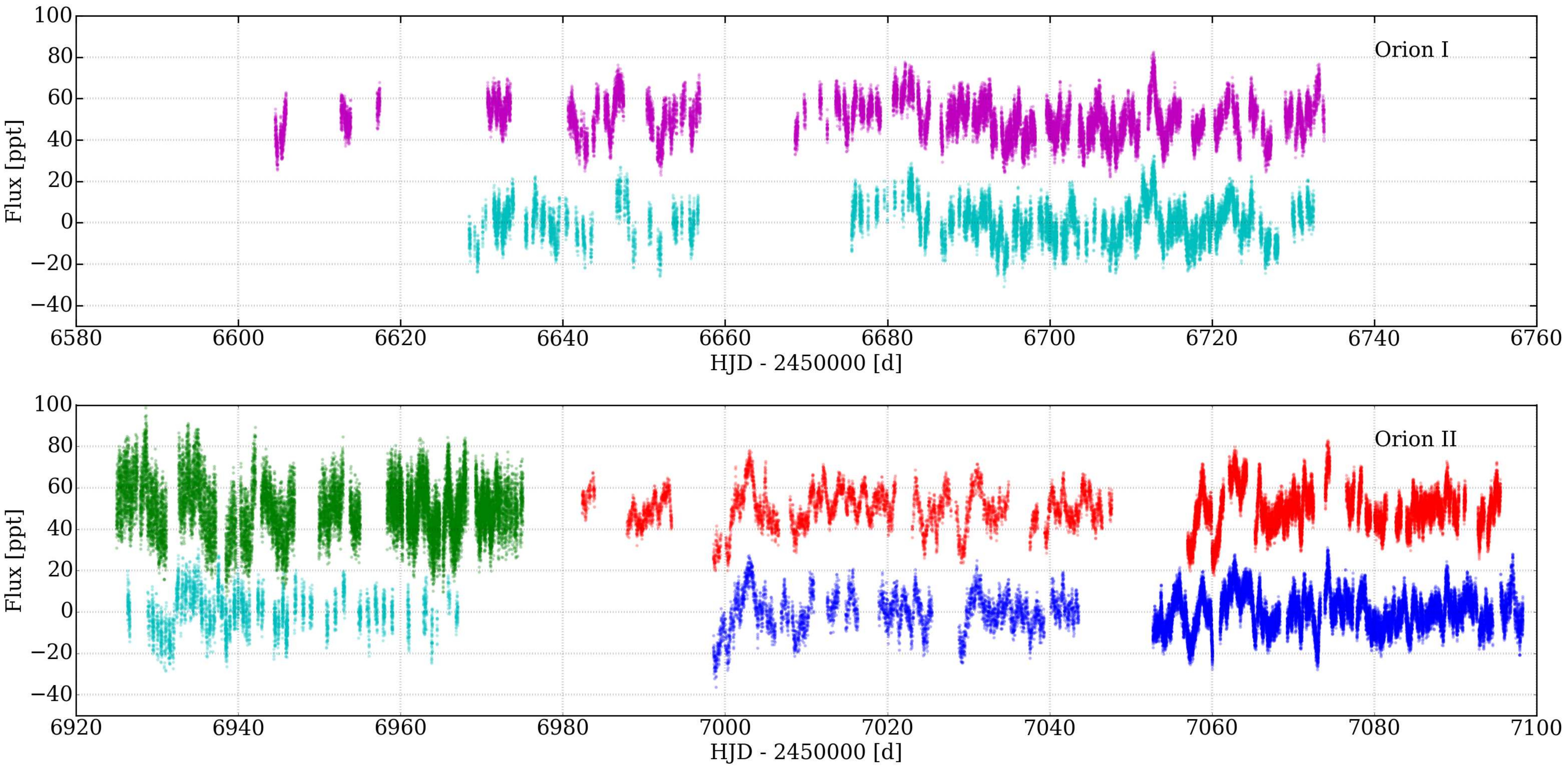}%
			\caption{All available BRITE photometry for \ZetOri, fully detrended and corrected for instrumental effects for the Orion\,I (\textit{top}) and Orion\,II (\textit{bottom}) observing campaigns. The color represents which nano-satellite of the BRITE-Constellation monitored \ZetOri: magenta for UBr, cyan for BAb, green for BTr, red for BHr, and blue for BLb. The flux variations are given in parts per thousand (ppt). Observations taken by a red nano-satellite (UBr, BTr, BHr) have an offset of 50\,ppt for increased visibility.}
			\label{fig:lightcurve_corrected}
\end{figure*}

\begin{figure*}[t]
		\centering
			\includegraphics[width=\textwidth, height = 0.66\textheight]{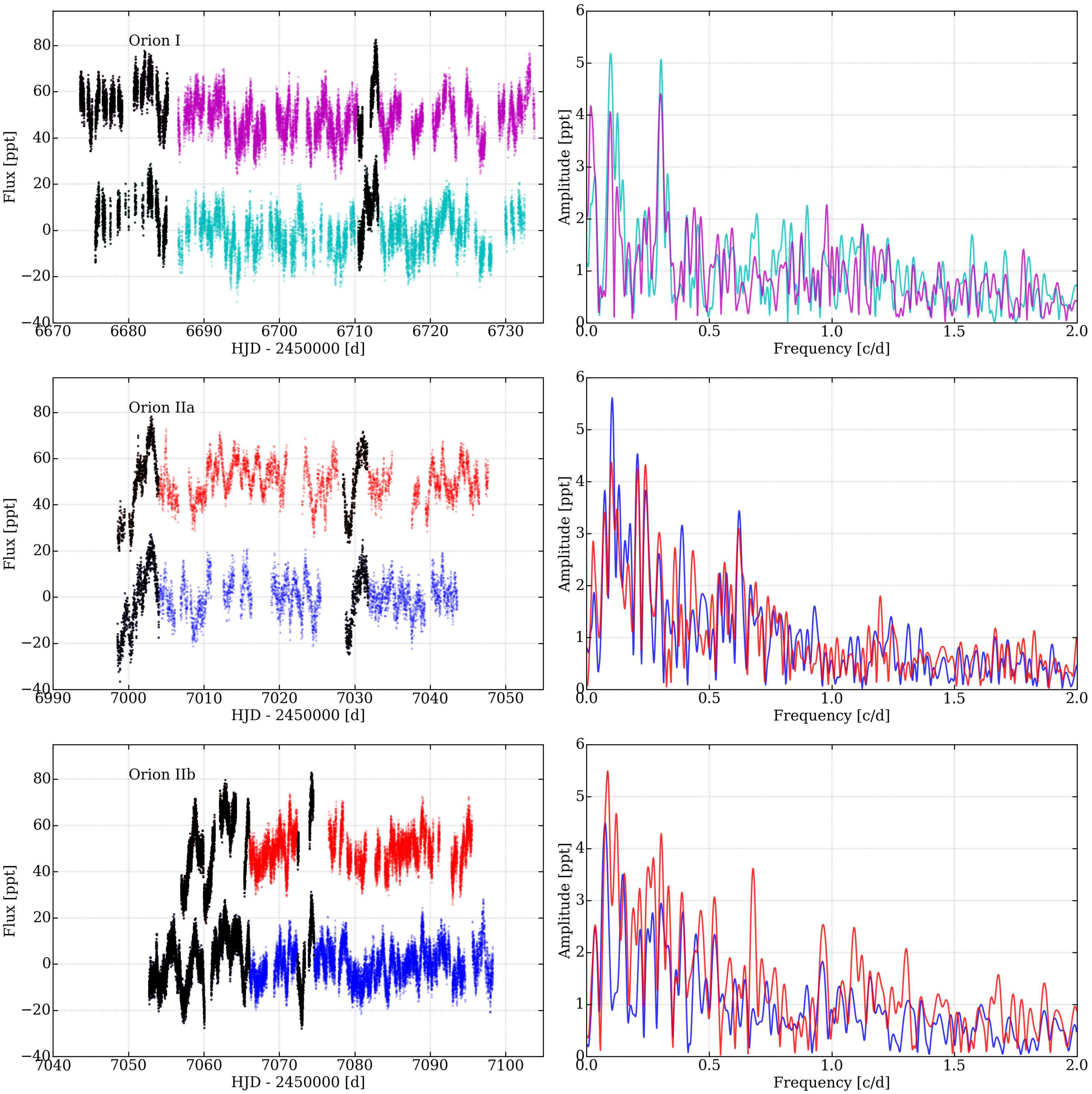}%
			\caption{\textit{Left}: BRITE photometry of \ZetOri during the three distinct epochs, used for the time series analysis. The color represents which nano-satellite of the BRITE-Constellation monitored \ZetOri: magenta for UBr, cyan for BAb, red for BHr, and blue for BLb. The regions having non-periodic events, which were excluded for a dedicated analysis, are marked in black (see text). \textit{Right:} Corresponding Lomb-Scargle periodograms of the full lightcurves. No significant periodic variability was found outside the region of 0 to 1\,\d. Variability is marked in parts per thousand (ppt).}
			\label{fig:lightcurve_studied}
\end{figure*}

\begin{table*}[t]
\caption{Extracted frequencies describing the photometric variability of \ZetOri seen in the studied BRITE lightcurves. The first column of each dataset presents the extracted frequencies for the full dataset, while the second column is for the non-periodic-event excluded photometry (these events are marked in black in Fig.\,\ref{fig:lightcurve_studied}). Uncertainties on the frequencies were determined using the Rayleigh criterion, and are thus very conservative. All frequencies are given in cycles per day.}
\centering
\tabcolsep=1pt
\begin{tabular}{p{2.5cm}||c||cccc||cccc||cccc}
\hline
\hline
& Combined & \multicolumn{4}{c||}{Orion\,I} & \multicolumn{4}{c||}{Orion\,IIa} & \multicolumn{4}{c}{Orion\,IIb}\\
Frequency &  & \multicolumn{2}{c}{Blue} & \multicolumn{2}{c||}{Red} & \multicolumn{2}{c}{Blue} & \multicolumn{2}{c||}{Red} & \multicolumn{2}{c}{Blue} & \multicolumn{2}{c}{Red}\\
$\delta f$	&	$\pm 0.003$	&$\pm	0.02	$&$\pm	0.02	$&$\pm	0.02	$&$\pm	0.02	$&$\pm	0.02	$&$\pm	0.03	$&$\pm	0.02	$&$\pm	0.02	$&$\pm	0.02	$&$\pm	0.03	$&$\pm	0.03	$&$\pm	0.03	$\\
\hline
$f_{\mathrm{rot}}$										&	0.147	&	0.15	&		&		&		&		&	0.14	&		&		&	0.15	&	0.16	&	0.16	&		\\
$2f_{\mathrm{rot}}$										&	0.292	&	0.30	&	0.30	&	0.30	&	0.30	&	0.30	&		&	0.30	&	0.30	&	0.31	&		&	0.30	&	0.31	\\
$3f_{\mathrm{rot}}$										&			&		&		&		&		&		&		&	0.43	&	0.44	&	0.45	&		&	0.47	&		\\
$4f_{\mathrm{rot}}$										&			&	0.58	&		&	0.60	&		&	0.62	&	0.62	&	0.62	&	0.61	&		&	0.59	&		&		\\
&&&&&&&&&&&&&\\																											
$f_{\mathrm{env}}$										&	0.100	&	0.10	&	0.09	&	0.10	&	0.09	&	0.10	&		&	0.10	&		&		&		&	0.09	&		\\
$2f_{\mathrm{env}}$										&			&		&		&		&	0.21	&	0.21	&		&	0.21	&		&	0.22	&	0.22	&		&		\\
$4f_{\mathrm{env}}$										&			&		&		&	0.41	&		&	0.38	&	0.37	&		&		&	0.39	&		&		&		\\
&&&&&&&&&&&&&\\																											
$f_{\mathrm{x}}$											&	0.036	&	0.03	&		&	0.03	&		&		&	0.04	&	0.03	&	0.02	&		&		&	0.03	&		\\
&&&&&&&&&&&&&\\																											
$f_{\mathrm{rot}} + f_{\mathrm{env}}$					&	0.248	&	0.24	&		&		&		&	0.23	&	0.24	&	0.24	&		&		&		&		&		\\
$3f_{\mathrm{rot}} + f_{\mathrm{env}}$					&			&		&		&		&		&	0.54	&	0.54	&	0.54	&		&	0.53	&		&		&		\\
$4f_{\mathrm{rot}} + f_{\mathrm{env}}$					&			&		&		&		&		&		&		&		&	0.68	&		&		&	0.68	&	0.68	\\
&&&&&&&&&&&&&\\		
$f_{\mathrm{env}} + f_{\mathrm{x}}$						&	0.118	&	0.12	&	0.12	&	0.13	&		&	0.13	&		&		&		&		&		&		&		\\
$f_{\mathrm{env}} + f_{\mathrm{x}}$						&	0.130	&		&		&		&		&		&		&		&		&		&		&		&		\\
$f_{\mathrm{env}} + f_{\mathrm{x}}$						&	0.135	&		&		&		&		&		&		&		&		&		&		&		&		\\
$f_{\mathrm{env}} - f_{\mathrm{x}}$						&	0.068	&		&		&		&		&		&		&	0.07	&	0.07	&	0.08	&	0.06	&	0.07	&	0.06	\\
&&&&&&&&&&&&&\\																											
$f_{\mathrm{rot}} - f_{\mathrm{x}}$						&	0.109	&		&		&		&		&		&		&		&		&		&		&		&		\\
$2f_{\mathrm{rot}} + f_{\mathrm{x}}$						&	0.320	&	0.33	&		&	0.34	&		&		&		&		&		&		&		&		&		\\
$2f_{\mathrm{rot}} - f_{\mathrm{x}}$						&	0.276	&		&		&		&		&		&		&	0.27	&	0.27	&	0.27	&		&	0.26	&		\\
\hline						
\end{tabular}
\label{tab:BRITE_frequencies}
\tablefoot{Most of the extracted frequencies are understood to be produced by three frequencies: $f_{\mathrm{rot}}$, corresponding to the stellar rotation, $f_{\mathrm{env}}$, related to the circumstellar environment; and $f_{x}$, a not-yet-understood frequency.}
\end{table*}

The BRITE-Constellation, an international collaboration between Austria, Canada, and Poland, currently comprises five fully functional and operational nano-satellites.  These BRITE nano-satellites aim to monitor stars brighter than $V\approx 5$\,mag using two colors, over various observing campaigns. These satellites each host a 3\,cm telescope, providing a wide field of view ($24\degr\times20\degr$) to simultaneously observe up to a few dozen stars \citep{2014PASP..126..573W}. Three nano-satellites perform their respective observations through a red photometric filter (550 -- 700\,nm; UniBRITE (UBr), BRITE Toronto (BTr), and BRITE Heweliusz (BHr)), while two employ a blue filter (390 -- 460\,nm; BRITE Austria (BAb), and BRITE Lem (BLb)). The various BRITE spacecraft have a low-Earth orbit, with orbital periods of about 100\,min ($\sim$0.07\,d). Therefore, they cannot continuously monitor targets, and instead, only observe them for a portion of the satellite orbit. The usable fraction of the orbit can be as large as 30\,\%, depending on the observing conditions, the satellite, and the studied field, before the telescope has to be pointed away \citep{2016PASP..128l5001P}.

Below, we present the BRITE observations acquired at two epochs, summarize the data reduction process, and describe the results.

\subsection{Observations}
\ZetOri was first observed by the two Austrian BRITE nano-satellites (BAb and UBr) as part of the commissioning field of the mission. This commissioning field is now known as the Orion\,I field.  In total, the observations lasted $\sim$130\,d, starting in early December 2013 and ending in mid March 2014: both BRITE nano-satellites studied the field with a high observing cadence whenever the satellites orbital phase permitted. The second visit by the BRITE-Constellation to \ZetOri occurred from late September 2014 until mid March 2015, lasting $\sim$170\,d in total. During this Orion\,II observing campaign, approximately the first 40\,\% of the observations were taken by BAb and BTr, and then the two Polish BRITE nano-satellites, BLb and BHr, took over. For some satellites, tests were performed to investigate whether onboard stackings of multiple, consecutive CCD images improved the data quality \citep{2016PASP..128l5001P}. This led to various observing cadences being adopted throughout the Orion\,II campaign, although each individual observation occurred with an exposure time of 1.0\,s. Table\,\ref{tab:BRITE_observations} lists the various BRITE observations of \ZetOri, taken during the Orion\,I and Orion\,II campaigns. The individual components of \ZetOri are not separately monitored by the BRITE-Constellation, since the pixel size ($27\,\arcsec$) is larger than the angular separation. Hence, the lightcurve contains all of its components, of which \ZetOri\,Aa is the brightest (see Table\,\ref{tab:zetori_params}).

To limit telemetry usage, no full CCD images are downloaded from the spacecrafts. Smaller CCD sub-rasters, with a size of $32\times32$\,pixels, are instead employed for each star. Once the observing campaign finishes, lightcurves are extracted from the CCD frames for all stars in the studied field, using circular apertures (Popowicz et al., in prep.). The raw lightcurve files have been corrected for intrapixel sensitivity, and provide additional metadata such as CCD centroid positions and CCD temperature. We then start from these datafiles to correct for any remaining instrumental effects.

\begin{figure*}[t]
		\centering
			\includegraphics[width=\textwidth, height = 0.66\textheight]{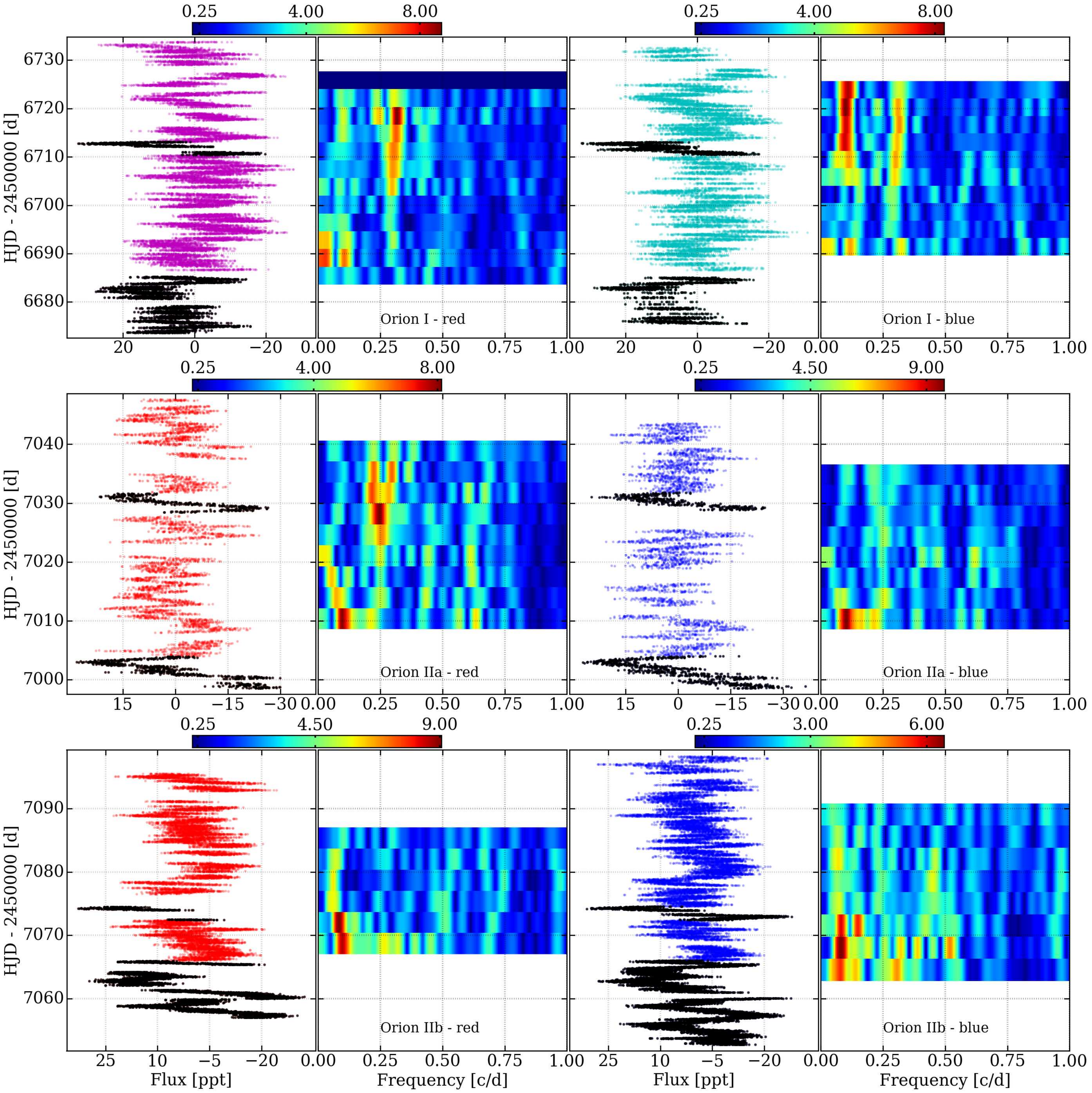}%
			\caption{Studied lightcurves of \ZetOri during the various epochs (\textit{first} and \textit{third} panels from left), compared to their respective STFTs (\textit{second} and \textit{fourth} panels).  Observations are indicated by a color representing the nano-satellite: magenta for UBr, cyan for BAb, red for BHr, and blue for BLb.  The amplitude of a given frequency in the sliding Lomb-Scargle periodogram is given by the color scale (in ppt; marked above each panel). The STFTs were calculated with a width of 20\,d, giving an effective frequency resolution of $\delta f = 1 / T = 0.05$\,\d, and a stepsize of 4\,d.}
			\label{fig:STFT}
\end{figure*}

\subsection{Data reduction summary}
The extracted BRITE photometry reveals the presence of several instrumental effects. Therefore, we constructed a data processing package to remove spurious signals and correct any remaining trends with instrumental variables. This package is now publicly available on Github\footnote{\url{http://github.com/bbuysschaert/BRITE_decor}}, and is based on our own experience with space-based photometry and the BRITE Data Analysis Cookbook\footnote{\url{http://brite.craq-astro.ca/doku.php?id=cookbook}} \citep[see also][]{2016A+A...588A..55P}. 

The first step of the reduction process consists of a conversion of the timing of the measurements to mid-exposure times.  This permits a clearer connection between BRITE lightcurves with different onboard stackings.  Next, we remove spurious signals, i.e., outliers, from both the metadata and the measured photometry, while accounting for their fluctuations with respect to time.  The metadata provided with the BRITE data consists of the onboard CCD temperature, $T_{\mathrm{CCD}}$, and the CCD centroid positions $x_c$ and $y_c$.  Once the data preparation is completed, we correct the BRITE photometry for the time- and temperature-dependent point spread function (PSF) by means of iterative B-spline fitting between flux and centroid positions within discrete subsets of the data based upon the temperature variability.  Finally, we account and correct for any significant correlation between the flux and $T_{\mathrm{CCD}}$, $x_c$, $y_c$, and $\phi_s$, the orbital phase of the satellite.  For a more detailed description of the various decorrelation steps, we refer to Appendix\,\ref{sec:appendix_correction}.

This data preparation and correction process was repeated for each data file provided. In addition to correcting for the instrumental signal, it also decreased the (instrumental) noise on the photometry, which we describe as the root mean square (RMS) of the flux, given as
\begin{equation}
\mathrm{RMS} = \sqrt{\frac{1}{N} \sum\limits_{i}^{N}{\frac{\sigma_i^2}{k_i}}}\ \mathrm{.}
\label{eq:RMS}
\end{equation}
\noindent Here, $\sigma_i$ and $k_i$ are the standard deviation of the flux and the number of observations within orbital passage $i$, respectively, and $N$ is the total number of orbital passages for the given setup file.

Finally, duty cycles for the different BRITE datasets were calculated, two of which have the most diagnostic value. One is the median duty cycle per satellite orbit, $\mathrm{D_{orb}}$, which indicates the fraction of the satellite orbit spent taking observations.  The other is $\mathrm{D_{sat}}$, which marks the portion of total satellite orbits used to successfully monitor the target.

The RMS values of the raw and the corrected photometry, as well as the duty cycles for the corrected lightcurves, are provided in Table\,\ref{tab:BRITE_observations}. The various corrected lightcurves are shown in Fig.\,\ref{fig:lightcurve_corrected}.  However, even after correction, differences remain in the data quality for \ZetOri between the various BRITE nano-satellites \citep{2016PASP..128l5001P}.

\subsection{Time series analysis of individual BRITE photometry}
\label{sec:results_timeseries}
Once all the BRITE observations were corrected, we adopted two different techniques to perform a time series analysis to study the periodic photometric variability of \ZetOri. First, we performed an iterative prewhitening approach to determine the periodicity of the dominant fluctuations \citep[see e.g.,][]{2009A+A...506..111D}. This method determines the most significant frequency in the periodogram for a dataset, fits a sinusoidal model with this frequency to the data, and determines the residual to the fit. We continue this approach iteratively, applying it to the residuals, determining new frequencies and updating previously extracted frequencies after each iteration until no significant variability remains. Second, we determined the short-time Fourier transform (STFT) of the datasets, calculating the periodogram within a moving window with a given width and stepsize. In this way, we studied the changes in both power and frequency of the periodic variability over time. 

\begin{figure*}[t]
		\centering
			\includegraphics[width=\textwidth, height = 0.66\textheight]{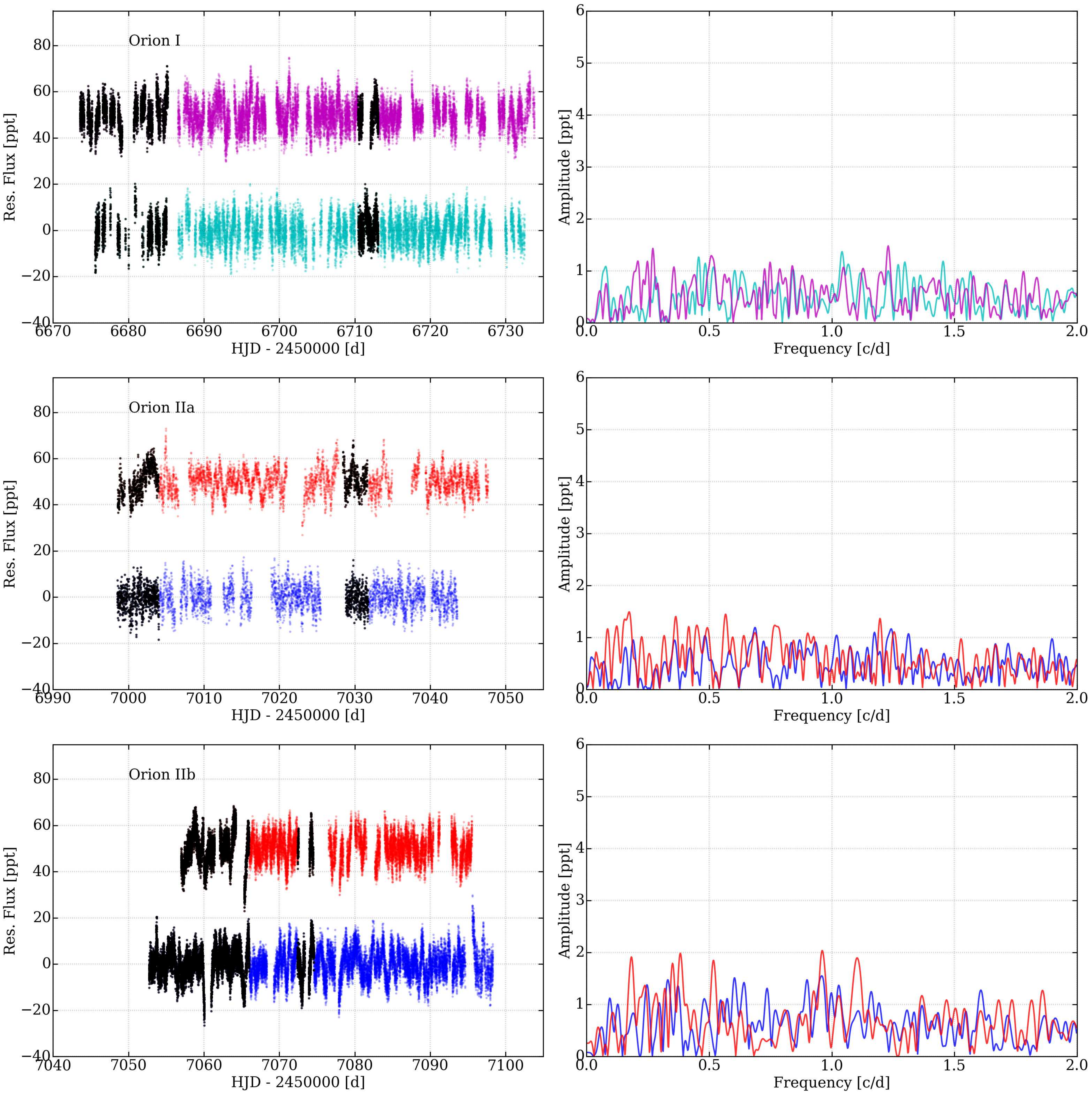}%
			\caption{\textit{Left}: Residual BRITE photometry of \ZetOri during three distinct epochs after all significant periodic variability was removed through iterative prewhitening. The same color-coding as in Fig.\,\ref{fig:lightcurve_studied} has been applied. \textit{Right:} Corresponding Lomb-Scargle periodograms of the residual lightcurves, represented with the same scale as in Fig.\,\ref{fig:lightcurve_studied}. The flux residuals contain some remaining variability, which is not significant in the Lomb-Scargle periodogram.}
			\label{fig:lightcurve_residual}
\end{figure*}

It became immediately apparent that the character of the variability of \ZetOri during the two BRITE observing campaigns is fairly different (see right panels of Fig.\,\ref{fig:lightcurve_studied}). Therefore, we studied the data from each campaign independently. In addition, we only considered the observations with the highest data quality (i.e., lowest RMS flux variability; see Table\,\ref{tab:BRITE_observations}) and the best duty cycle for the time series analysis. The considered Orion\,II data were further subdivided into two independent sets, because of the inconsistent number of onboard stacked images, which caused issues in the weighting of the frequency analysis, and because of differences in the observed variability. Thus, we investigated data for three different epochs, namely Orion\,I, Orion\,IIa, and Orion\,IIb. For each epoch, we have a red and a blue photometric dataset of almost equal length, which agree quite well once corrected for the instrumental effects. We show these six investigated lightcurves in Fig.\,\ref{fig:lightcurve_studied}, with their corresponding Lomb-Scargle periodograms indicating their periodic variability.

Each individual BRITE lightcurve underwent separate iterative prewhitening to determine the significant periodic variability, using a Lomb-Scargle periodogram with an oversampling factor of 10 \citep{1976Ap+SS..39..447L, 1982ApJ...263..835S}. We calculated the significance of a frequency peak in the periodiogram using a signal-to-noise ratio (S/N) criterion. We considered it significant when a S/N larger than four was reached \citep[][]{1993A+A...271..482B}, within a frequency window of 4.0\,\d. Such a large window was chosen because all detected variability is present in the low-frequency regime ($<\,2.0$\,\d) of the periodogram; therefore, the full window is not effectively used for the S/N calculation. Finally, only the frequency domain below 10\,\d was considered for investigation because the satellite orbital frequency produces a forest of very strong alias peaks every 14.3 -- 14.8\,\d; the exact value depends on which the nano-satellite is used.

The extracted frequencies from the various lightcurves were examined via a pairwise comparison between the two colors of the same epoch, and also over the three observing epochs. When we did not retrieve the same frequency in at least two datasets, within the confidence interval, that frequency was discarded for the model lightcurve. We list the final set of extracted frequencies for each of the six datasets in Table\,\ref{tab:BRITE_frequencies}. Because of the moderate duty cycle and because of the presence of gaps within the BRITE observations, we used the Rayleigh criterion for the confidence interval of the extracted frequencies. The conservative uncertainty for a frequency is thus $\delta f = 1 / T$, where $T$ is the total time span of the data.

While investigating the individual BRITE lightcurves, the phase-folded photometry with the dominant periodicities, and the residual lightcurves of the detailed frequency analysis, we identified peculiar time regions in the lightcurves. These regions stand out through their pronounced deviations to the iterative prewhitening model, as the observed brightness differs substantially from the model for the periodic variations.  As such, we assumed that these time regions are influenced more strongly by non-periodic or non-cyclic variability than the remainder of the BRITE photometry.  To define these time regions more accurately, we used an iterative approach to study the simultaneous blue and red residual lightcurves to the iterative prewhitening model.  These new lightcurves were then subjected to a new detailed frequency analysis.  We continued to exclude observations until no obvious deviations from the periodic variability model were noted.  The final rejected time domains are marked in black in Fig.\,\ref{fig:lightcurve_studied}.  An independent frequency analysis was then performed on these alternative lightcurves, for which we provide the deduced periodic variability in Table\,\ref{tab:BRITE_frequencies}.  The particular regions of more pronounced non-periodic variability are further discussed in Sect.\,\ref{sec:discussion_circst}.

In total, we extracted 16 different frequencies in the frequency domain 0 -- 1\,\d (see Table\,\ref{tab:BRITE_frequencies}). The majority of these 16 frequencies are understood to originate from three independent frequencies, which construct through harmonics and simple frequency combinations most of the extracted periodic variability. The three independent frequencies are $f_{\mathrm{rot}}\,=\,0.15\pm0.02$\,\d, $f_{\mathrm{env}}\,=\,0.10\pm0.02$\,\d, and $f_{\mathrm{x}}\,=\,0.04\pm0.02$\,\d. We append this reasoning for each frequency in Table\,\ref{tab:BRITE_frequencies}. All of these fundamental frequencies, and some particular ones, are discussed in more detail in Sects.\,\ref{sec:spectroscopy} and \ref{sec:discussion}.

The variations with the largest detected amplitudes correspond to $2f_{\mathrm{rot}}$ and $f_{\mathrm{env}}$ (see Table\,\ref{tab:BRITE_amplitudes}), and are clearly distinguishable in the right panel of Fig.\,\ref{fig:lightcurve_studied}.  Differences in amplitudes are noted between the various epochs.  In particular $4f_{\mathrm{rot}}$, which is related to (the photospheric origin of the) DACs, is stronger in the red data.  However, differences are likely related to the non-periodic variability, since amplitudes are altered when the regions strongly affected by the non-periodic variability are excluded (marked in black in Fig.\,\ref{fig:lightcurve_studied}).  Moreover, owing to the considerable amount of corrections applied, we expect our formal uncertainty on the determined amplitude to be strongly underestimated.  In addition, different responses between the various nano-satellites could occur.  Hence, we refrain and warn against overinterpreting the possible differences in amplitude between the simultaneous red and blue photometry.

For each individual BRITE lightcurve we also investigated the STFT. To be able to study the low-frequency regime of these periodograms, a sufficient frequency resolution needed to be achieved. Hence, the chosen time window spanned 20\,days, corresponding to a frequency resolution of 0.05\,\d. The stepsize was set to be 4\,d, because if it were too small it would smear out variability over time, while if too large it would hamper the interpretation of the observed variations. If the $\mathrm{D_{sat}}$ duty cycle of the data within the time window did not reach 50\,\%, no periodogram was determined for that step. The calculated periodograms were Lomb-Scargle periodograms with an oversampling factor of 10.  We show the final STFTs in Fig.\,\ref{fig:STFT}, organized per observing epoch, including the regions marked in black in Fig.\,\ref{fig:lightcurve_studied}.

While the constraint on the duty cycle, $\mathrm{D_{sat}}$, inhibits the determination of a Lomb-Scargle periodogram at each step, the STFT analysis is still able to confirm the differences between each studied epoch of BRITE photometry. These differences are compatible with the results from the iterative prewhitening approach. Moreover, within each epoch, the amplitude of prominent frequencies varies with respect to time, and often increases whenever regions with strong non-periodic variability occurred (marked in black in Fig.\,\ref{fig:lightcurve_studied}), indicating the presence of non-periodic photometric variability for \ZetOri. Typically, the most pronounced variability of the iterative prewhitening analysis can be linked to certain regions of the dataset, with $2f_{\mathrm{rot}}$ the general exception, as indicated by the STFTs (see Fig.\,\ref{fig:STFT}). Minor shifts in frequency for certain variability can be noted with respect to time, yet still fall within the determined frequency resolution dictated by the adopted time window of the STFT (0.05\,\d).

Once all periodic variability corresponding to the significant frequencies was extracted from the BRITE photometry of \ZetOri using the iterative prewhitening approach, we again calculated the residuals to this model. Although the periodogram of these residuals did not indicate any remaining significant periodic variability, the residual flux did still contain visible variability (see Fig.\,\ref{fig:lightcurve_residual}). When comparing the residuals of each lightcurve to the version of the same lightcurve with the earlier discarded regions, we noted that more variability remains within the discarded regions in the residuals of the original lightcurve.  This, again, confirmed the need for the separate analysis, where the strongest non-periodic (and non-cyclic) variability was excluded.

\begin{figure}[t]
		\centering
			\includegraphics[width=\columnwidth, height = 0.33\textheight]{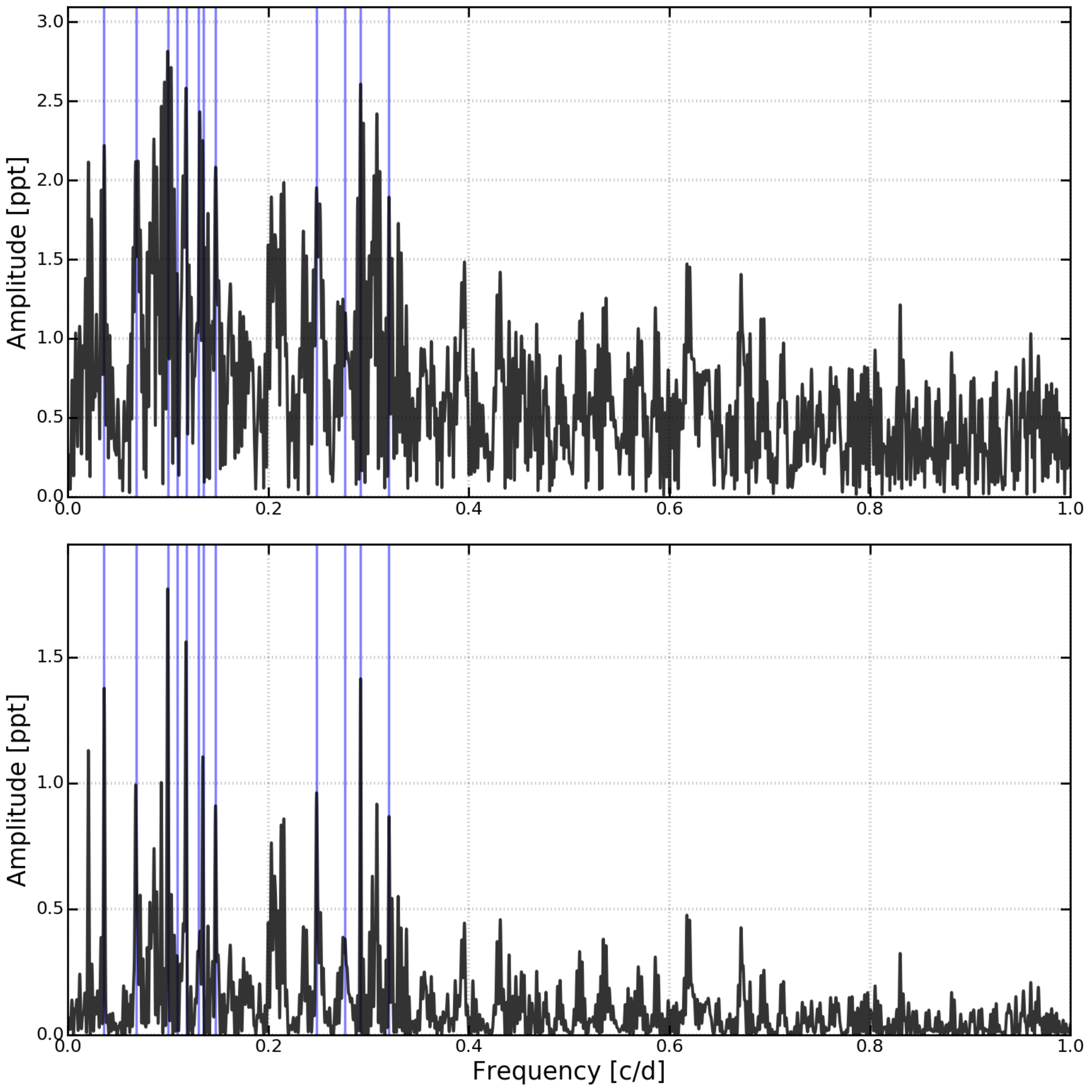}%
			\caption{Different periodograms of the combined BRITE lightcurve for \ZetOri. \textit{Top}: Regular Lomb-Scargle periodogram, significantly influenced by the spectral window. \textit{Bottom}: CLEAN periodogram, using ten iterations and a gain of 0.5, showing the variability more clearly.  Extracted significant periodic variability is marked in blue.}
			\label{fig:lightcurvecombined_periodogram}
\end{figure}

\subsection{Time series analysis of combined BRITE photometry}
\label{sec:results_timeseries_full}
Aiming to increase the frequency resolution and decrease the noise level, we merged various BRITE observations to one long-baseline lightcurve. This was only possible with the increased comprehension of the individual BRITE photometry for \ZetOri. Studying the duty cycles and data quality of the respective BRITE lightcurves, we opted to include all red and blue data from all three epochs, i.e., Orion\,I, Orion\,IIa, and Orion\,IIb, as well as the BRITE Toronto-1 data (see Table\,\ref{tab:BRITE_observations}). As a first step, we averaged and rebinned the observations to one single measurement per satellite orbit passage, permitting an easier connection between datasets with a different number of onboard stackings. Next, we merged all these rebinned data, ignoring the color information from the observations. Although minor differences between the two colors are present, as we show later, the individually studied lightcurves agree rather well. Moreover, we increased the duty cycle of the full lightcurve significantly by combining the simultaneous lightcurves, since minor differences for the orbital period of the BRITE nano-satellites lead to observations being taken at different times. This new, combined lightcurve spans 493.7\,days, over both the Orion\,I and Orion\,II BRITE observing campaigns, with a 191.2\,day time gap between the two parts.

Similar to the study of the individual BRITE datasets, we performed a times series analysis to study the photometric variability of \ZetOri, using an iterative prewhitening approach on a Lomb-Scargle periodogram with an oversampling factor of 10. Again, we employed the S/N criterion to determine the significance of a given frequency, using a frequency window of 2\,\d, and only investigated the frequency domain below 10\,\d. We show this periodogram in Fig.\,\ref{fig:lightcurvecombined_periodogram}; we also show a CLEAN periodogram \citep[][]{1987AJ.....93..968R}, because the large time gap in the dataset influences the spectral window of the periodogram considerably. In total, we determined and extracted 12 significant frequencies, which are given in Table\,\ref{tab:BRITE_frequencies}. Because of the large time gap present in the combined photometry, the Rayleigh criterion might be a too optimistic value for the frequency uncertainty. Therefore, we only used the time span when BRITE observations for \ZetOri\,Aa were actually taken (about 300\,days).  This leads to an altered frequency resolution of $\delta f = 1 / T_{\rm obs} = 0.003$\,\d.

Several of the 12 retrieved frequencies agree with values we determined from the individual data, albeit determined with a higher precision. However, some of the previously determined frequencies were not recovered, e.g., 0.60\,\d and 0.68\,\d, since they appear to be much more prominent in the red data of Orion\,IIb, and fairly weak in the Orion\,I photometry. Thus, these became less significant when we combined the red and blue BRITE photometry of the various epochs.

\section{CTIO 1.5\,m/CHIRON \'{e}chelle spectroscopy}
\label{sec:spectroscopy}
\subsection{Observations}
\label{sec:obs_spectroscopy}

\begin{figure}[t]
		\centering
			\includegraphics[width=\columnwidth, height = 0.66\textheight]{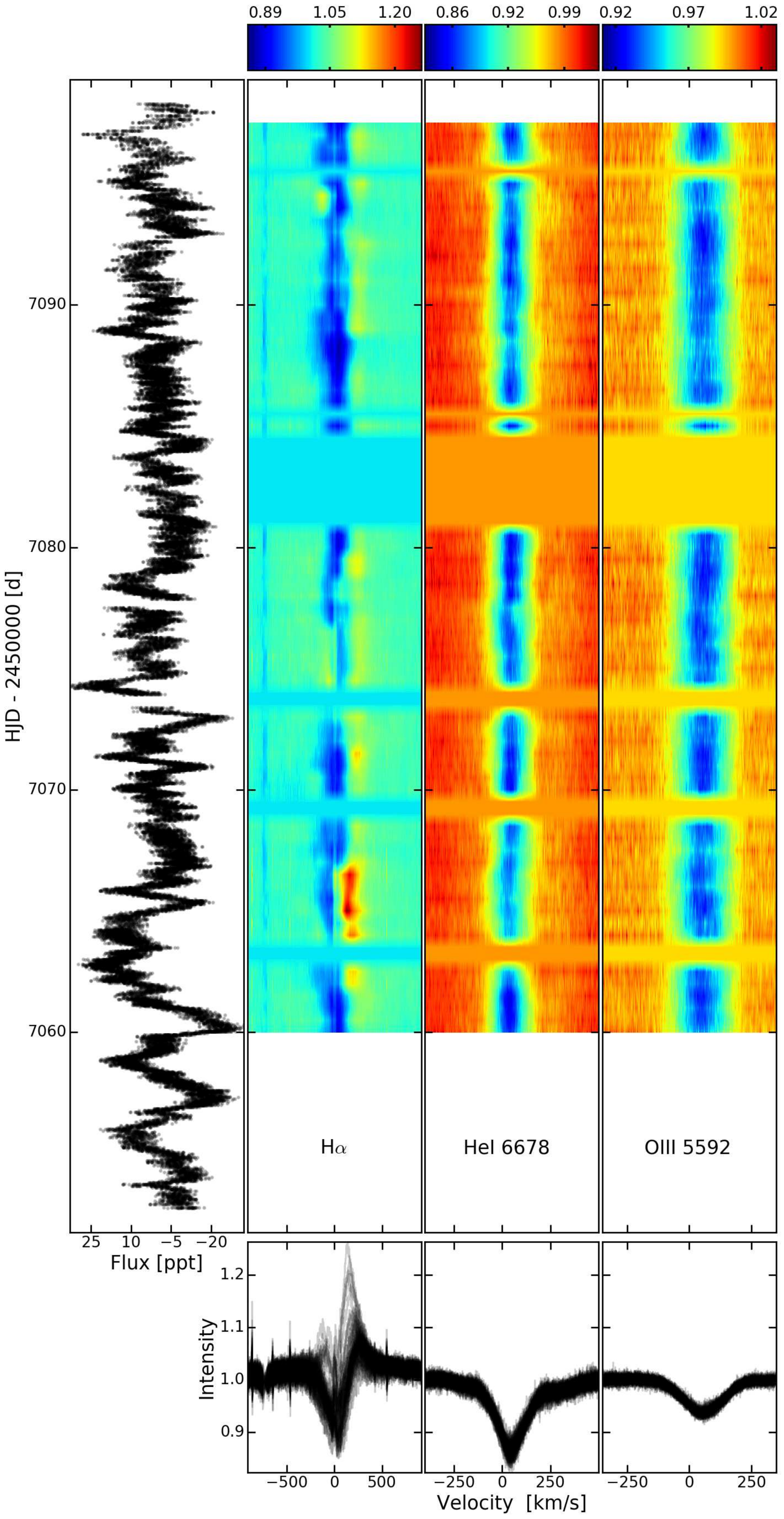}%
			\caption{Comparison between the simultaneous CHIRON spectroscopy and the BRITE photometry during the Orion\,IIb campaign. \textit{Left}: Superimposed red and blue BRITE photometry during the Orion\,IIb campaign. \textit{Right}: Dynamical representation of different lines from the CHIRON spectroscopy taken simultaneously with the Orion\,IIb BRITE campaign. \textit{Bottom}: Overlayed profiles of the spectroscopic lines.}
			\label{fig:CHIRON_dynamical}
\end{figure}

\begin{figure}[t]
		\centering
			\includegraphics[width=\columnwidth, height = 0.33\textheight]{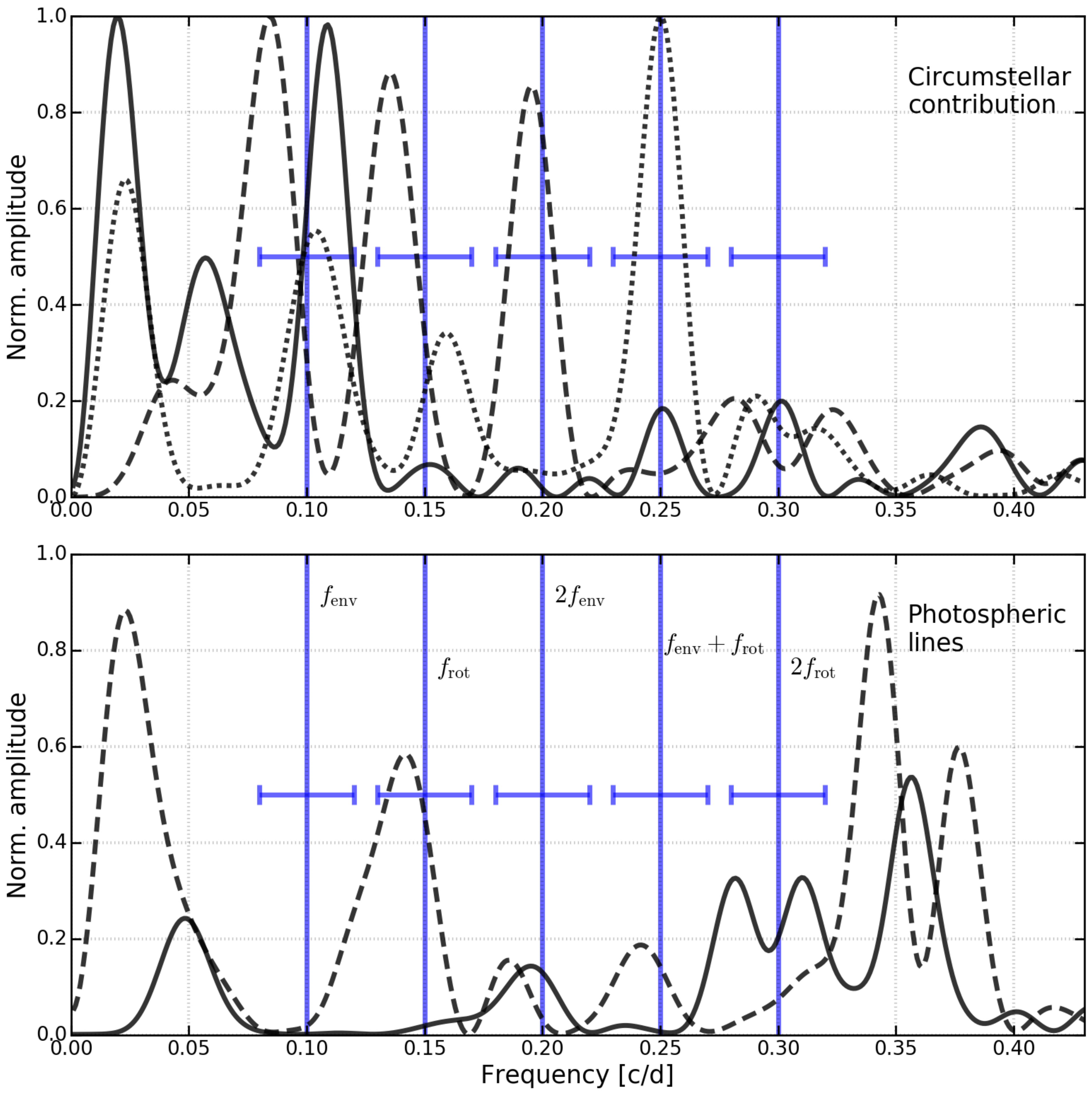}%
			\caption{CLEAN periodograms for EWs of various spectroscopic lines. \textit{Top}: Periodograms for lines being influenced by the circumstellar environment (solid - H$\alpha$; dashed - \ion{He}{I}\,$\lambda\lambda6678.2$; dotted - \ion{He}{I}\,$\lambda\lambda4921.9$). \textit{Bottom}: Periodograms for purely photospheric lines (solid - \ion{O}{III}\,$\lambda\lambda5592.3$; dashed - \ion{C}{IV}\,$\lambda\lambda5801.3$). Dominant frequencies found in the individual BRITE photometry are marked in blue (from left to right: $f_{\mathrm{env}}$, $f_{\mathrm{rot}}$, $2f_{\mathrm{env}}$, $f_{\mathrm{env}} + f_{\mathrm{rot}}$, and $2f_{\mathrm{rot}}$).}
			\label{fig:CHIRON_periodogram}
\end{figure}

\ZetOri\,A was observed 60 times between 2015 February 6 and 2015 March 16 (typically twice per night) with the CHIRON \'{e}chelle spectrograph mounted on the CTIO 1.5-m telescope \citep{2013PASP..125.1336T}. This fiber-fed spectrograph was used in slicer mode, yielding an effective resolving power of $R \sim 80000$, and covers the region from 459 to 760\,nm. All spectra were corrected for bias and flat-field effects and wavelength calibrated through the CHIRON pipeline. Each visit consisted of two consecutive spectra with an individual exposure time of 8 seconds. These extracted spectra were combined into a single spectrum, resulting in individual visits having a S/N per resolving element of at least 75 in the continuum near H$\alpha$. The spectra were blaze-corrected and normalized to a unit continuum.  Several telluric features are present in the H$\alpha$ region. To remove these lines, we used a telluric template spectrum convolved to the instrumental response in the region surrounding H$\alpha$, as measured from the FWHM of isolated H$_2$O lines. The depth of the template lines was adjusted interactively until the residual spectrum was smooth.  A log of the different spectroscopic observations is given in Table\,\ref{tab:CHIRON_log}.

The fiber going to CHIRON has a diameter of 2.7" on the sky, which is sufficiently small to exclude \ZetOri\,B, which is currently $\sim2.4$" away.  Therefore, the CHIRON data only contain spectral features of \ZetOri\,Aa and \ZetOri\,Ab, albeit with different contributions.  The short time span and especially the quality of the CHIRON data leads to a very uncertain spectral disentangling.  Including archival spectroscopy to cover the full (inner) orbit would most likely only add to that uncertainty because of differences in the optical design of the instruments.  In the remainder of this paper, we consider that any spectral variability in the studied frequency regime originates from the magnetic supergiant \ZetOri\,Aa, first because the short time span of the CHIRON data does not contain significant radial velocity variations from the binary motions, and because \ZetOri\,Ab is fainter, leading to a respective contribution of at most 10\,\% \citep[as noted from earlier spectral disentangling attempts by][]{2015A+A...582A.110B}, so that any mild spectral variability in \ZetOri\,Ab will have an effect comparable to the noise level.

In the following, we discriminated between the different spectral lines of \ZetOri\,Aa on their (partial) origin. First, we present the study and results for lines containing a contribution from the circumstellar environment or the stellar wind (Sect.\,\ref{sec:results_spectroscopy_circumst}) and thenn we do the likewise for purely photospheric lines (Sect.\,\ref{sec:results_spectroscopy_photosph}).

\subsection{Analysis of the H$\alpha$ spectra and \ion{He}{I} lines}
\label{sec:results_spectroscopy_circumst}

We limited the investigation of the spectral lines with a circumstellar contribution to H$\alpha$, which is the Balmer line most influenced by the circumstellar environment, and two of the strongest \ion{He}{I} lines, i.e., \ion{He}{I}\,$\lambda\lambda6678.2$ and \ion{He}{I}\,$\lambda\lambda4921.9$.

In Fig.\,\ref{fig:CHIRON_dynamical}, we show the dynamical representation of the H$\alpha$ and \ion{He}{I}\,$\lambda\lambda6678.2$ lines and compare them to the simultaneous BRITE photometry of the Orion\,IIb epoch. As expected for a massive star, these lines vary with time, and the variation is most pronounced for the P\,Cygni profile of H$\alpha$. To study the periodic component of this variability, we measured the net equivalent width (EW) of the lines. For H$\alpha$ we integrated over both the absorption and emission features. The integration limits for the measurements spanned from $v=-1000$ to $1000$\,km\,s$^{-1}$ for H$\alpha$, and was typically from $v=-500$ to $500$\,km\,s$^{-1}$ for the \ion{He}{I} lines, with $v$ being the radial velocity. Next, we calculated and investigated the periodograms of the EWs to determine the most dominant periodic behavior. Since the data were taken from Earth and have limited coverage, they have an unfavorable spectral window. Therefore, we resorted to CLEAN periodograms, since these limit the influence by the spectral window, using ten iterations, a gain of 0.5, and ten times oversampling. Only a limited frequency regime from 0 to 0.5\,\d was investigated, because of the presence of very strong 1\,\d aliasing effects and a strong reflection effects. These normalized periodograms of the EW are shown in the top panel of Fig.\,\ref{fig:CHIRON_periodogram}.

The data did not permit a detailed frequency extraction and analysis. Instead, we compared the periodogram with that of the BRITE photometry to determine which known frequencies are needed to explain the variability of the spectral lines. For all lines, we retrieved prominent variability at $0.09 - 0.11 \pm 0.03$\,\d, corresponding to $f_{\mathrm{env}}$. In addition, the EWs of the \ion{He}{I} lines also exhibit fluctuations interpreted as the rotation at $f_{\mathrm{rot}}$, while we only marginally retrieve this variability for H$\alpha$. Different combinations between independent frequencies are again observed, similar to the BRITE photometry.  We refrain from further interpreting the power below 0.05\,\d in Fig.\,\ref{fig:CHIRON_periodogram}, because additional testing showed that only these amplitudes are severely altered by the upper frequency limit of the CLEAN routine.  Increasing this upper limit decreases the power below 0.05\,\d, by moving it towards the $\sim 1$\,\d area, indicating the uncertain origin and nature of this variability and possibly influenced by the CLEAN routines.   Finally, the power around $0.32 - 0.38$\,\d should be treated with care, since it is likely a combination of reflections of power in the $0.60 - 0.68$\,\d region around the 1\,\d alias, and variations with $2f_{\mathrm{rot}}$.

\subsection{Analysis of photospheric lines}
\label{sec:results_spectroscopy_photosph}
Following \citet{2008MNRAS.389...75B} and \citet{2015A+A...574A.142M}, we selected \ion{O}{III}\,$\lambda\lambda5592.3$ and \ion{C}{IV}\,$\lambda\lambda5801.3$ as photospheric absorption lines for further investigation, because they are not influenced (or only weakly) by the stellar winds and the circumstellar environment. The same method as in Sect.\,\ref{sec:results_spectroscopy_circumst} was used to study the periodic variations of the EW of these lines, which was typically determined in the velocity region spanning $v=-400$ to $400$\,km\,s$^{-1}$. We show the normalized determined CLEAN periodograms for these lines in the bottom panel of Fig.\,\ref{fig:CHIRON_periodogram}.

The most striking result for the periodograms of the purely photospheric lines is the strong difference with their corresponding periodograms of the H$\alpha$ and \ion{He}{I} lines. Indeed, the frequency region around $f_{\mathrm{env}}$ appears to be absent of any significant power for the purely photospheric lines. Instead, the strongest power is present below $0.05$\,\d and in the  $0.32 - 0.38$\,\d region, which should be treated with care, similar to the periodograms for the lines with a circumstellar contribution.  Some power around $f_{\mathrm{rot}}$ is also present for the \ion{C}{IV} line.

\section{Discussion}
\label{sec:discussion}
According to the V magnitudes determined by \citet{2013A+A...554A..52H} for all three components, and their similar temperature of about 29000\,K, we deduce that 77\,\% of the received flux originates from \ZetOri\,Aa, 10\,\% from \ZetOri\,Ab, and 13\,\% from \ZetOri\,B.  These values are compatible with the contribution of \ZetOri\,Ab to the Narval data, as determined from the spectral disentangling by \citet{2015A+A...582A.110B}.  Therefore, throughout the discussion of the extracted frequencies from the BRITE photometry, we expect that all detected variability is caused by the magnetic supergiant \ZetOri\,Aa and its circumstellar environment.  In the same way, variations observed in the CHIRON spectroscopy are likely related to \ZetOri\,Aa (see Sect.\,\ref{sec:obs_spectroscopy}). Moreover, the possible high-degree $\beta$\,Cep pulsations of the B component \citep{2013A+A...554A..52H} would fall in a different frequency regime ($>2$\,\d) than that of the currently determined periodic variability.  In addition, high-degree pulsation modes are not often visible in photometry due to partial cancellation effects.  Yet, it cannot be fully excluded that strong variability in \ZetOri\,Ab or \ZetOri\,B could produce weak frequency peaks in the data presented here.

\subsection{Rotation}
\label{sec:discussion_rotation}
In each and every studied BRITE lightcurve, we determined variability related to half the literature rotation period, corresponding to a frequency of $0.30\pm 0.02$\,\d (see Table\,\ref{tab:BRITE_frequencies}). However, the amplitude corresponding to this variation differs from one epoch to the next, most likely due to non-periodic photometric events happening in the lightcurve altering the determined amplitudes of the periodic variability. Excluding the regions of the lightcurve where the non-periodic signals occur the strongest has no considerable effect on the extraction of this frequency: we still recover it for four out of six lightcurves. The only exception is the blue Orion\,II data. Using the combined BRITE photometry, we retrieve a similar frequency, $0.292\pm 0.003$\,\d.

The presence of photometric variability with exactly half the rotation period is often observed for hot or massive magnetic stars \citep[e.g., $\sigma$\,Ori\,E;][]{2015MNRAS.451.2015O}, and is related to the dipolar structure of the large-scale magnetic field, coming into and leaving the observer's line of sight. Indeed, the conditions at the magnetic poles are fairly different from the rest of the star, both at the stellar surface and in its very close circumstellar proximity.  It was not possible to accurately verify the phase difference between the rotational modulation of the BRITE photometry of \ZetOri\,Aa and the longitudinal field measurements of \citet{2015A+A...582A.110B} because of the substantial errorbar on the derived rotation period.  Nevertheless, since our derived rotational frequency and the published frequency from the magnetic analysis are compatible, we consider it likely that the variability is related to the poles of the dipolar, fossil magnetic field of \ZetOri\,Aa, as seen in many other magnetic massive stars.

In addition to the frequency corresponding to half of the rotation period, we also extracted $f_{\mathrm{rot}}$, $3f_{\mathrm{rot}}$, or $4f_{\mathrm{rot}}$ for some of the BRITE lightcurves. Of these, the anomalous variability is related to $3f_{\mathrm{rot}}$, as it would indicate the presence of three surface spots. Such a geometrical magnetic configuration is considered highly unlikely. Yet, a strong bias to favor a small number of (equidistant) spots exists: the effect of more than a few individual spots on the stellar surface of \ZetOri\,Aa on the lightcurve would be merged, effectively masking their presence and mimicking the variability of only a few spots. However, detailed spot modeling is currently beyond the scope of this work, since it requires a higher precision of $P_{\mathrm{rot}}$ that is not achievable with the current dataset. We discuss the fourth harmonic in more detail in Sect.\,\ref{sec:discussion_DAC}, as it is likely related to DACs.

Combining measurements of the frequencies believed to be $f_{\mathrm{rot}}$ and its higher harmonics from the individual BRITE photometry to increase the precision, we obtain $P_{\mathrm{rot}} = 6.65\pm0.28$\,d. Doing this for the combined BRITE photometry, yields $P_{\mathrm{rot}} = 6.82\pm0.19$\,d. These values are consistent within the error bars with the rotation period of \citet[][i.e., $6.83\pm0.08$\,d]{2015A+A...582A.110B}. A more precise measurement will be difficult without a detailed understanding of the variability of the supergiant \ZetOri\,Aa, in particular, of its circumstellar environment.

Because of the limited temporal resolution of the spectroscopic dataset, we could not perform a detailed frequency analysis of the spectroscopic variability to accurately determine the rotation period from spectroscopy.

\subsection{DACs}
\label{sec:discussion_DAC}
Most of the studied BRITE photometry indicates that \ZetOri\,Aa is variable with $P_{\mathrm{rot}}/4$.  This periodicity coincides with the DAC recurrence timescale derived by \citet{1999A+A...344..231K} for \ZetOri\,Aa.  Indeed, our derived value of $\mathrm{t_{rec}} = 1.67\pm0.06$\,d agrees with the literature value of $1.6\pm0.2$\,d (i.e., $\mathrm{f_{rec}} = 0.625\pm0.075$\,\d). During our analysis, both the STFT and the iterative prewhitening of the individual BRITE lightcurves indicated that the amplitude of this frequency varies with time (e.g., Table\,\ref{tab:BRITE_amplitudes} and Fig.\,\ref{fig:STFT}). Under the assumption that DACs are related to surface inhomogeneities, the description of four (transient) enhanced brightness regions evenly distributed in longitude over the stellar surface is the most likely. However, with the current knowledge of this object, we were unable to corroborate or disprove the exact mechanism that creates the hypothetical enhanced brightness regions leading to the CIRs.  Higher quality observations, having a high duty cycle and long baseline, are needed to explore the origin of these inhomogeneities and study their stability over time.

Variability with a similar timescale seems to be present in the studied spectroscopic lines. Yet, the limited cadence and poor spectral window, mainly related to reflection effects over 1\,\d, did not permit us to exclude a purely instrumental origin.

Finally, red Orion\,IIb data show evidence of photometric brightness variations with a frequency of $0.68\pm0.02$\,\d. We consider this power to be related to the four surface spots, yet slightly distorted to a higher frequency because of a frequency combination with $f_{\mathrm{env}}$. This timescale also matches, within the uncertainty, the Hipparcos periodicity of 1.407\,d discussed by \citet{2004MNRAS.351..552M}.

\subsection{Circumstellar environment}
\label{sec:discussion_circst}
Comparing the variability of the CHIRON spectroscopy with that of the BRITE photometry results in a particular agreement and discrepancy. Both the BRITE photometry and spectroscopic lines with a circumstellar contribution (i.e., H$\alpha$ and \ion{He}{I} lines) show variability with $f_{\mathrm{env}} = 0.100 \pm 0.003$\,\d, while it is strikingly absent in the photospheric lines. Therefore, we conclude that the BRITE photometry contains a contribution from the circumstellar environment of \ZetOri\,Aa, and that this environment is periodically (or possibly cyclically) variable with $f_{\mathrm{env}}$, and additionally indicates non-periodic variability. Studying the individual BRITE lightcurves without these strong non-periodic events, we no longer recover this frequency for the Orion\,IIa and Orion\,IIb data.  We therefore consider these parts of the BRITE photometry to be dominated by non-periodic variability originating at the circumstellar environment. Figure\,\ref{fig:CHIRON_dynamical} indeed shows that these regions of the lightcurve are correlated with strong changes in shape and power of the P\,Cygni profile of the H$\alpha$ line, which probes the variability within the circumstellar environment.

We deem it therefore probable that the circumstellar environment causes the observed differences in variability between the three BRITE epochs, either by decreasing the detected flux by obscuring parts of the stellar surface or temporarily increasing the brightness. An example of the effect of such an increase in brightness can be seen in the Orion\,I data, where the rotational variability with $2f_{\mathrm{rot}}$, coming from \ZetOri\,Aa itself, is still visible during the event (i.e., top left panel of Fig.\,\ref{fig:lightcurve_studied} at HJD = 2456714).

Since \ZetOri\,Aa hosts a large-scale magnetic field at the stellar surface, the circumstellar environment would also be influenced by it.  Previous studies have shown that the dynamical magnetosphere of \ZetOri\,Aa is rather weak \citep{2015A+A...582A.110B}.  We do not find strong evidence for rotational variability in H$\alpha$ compatible with a very weak or absent magnetosphere.  If the magnetosphere were causing periodic photometric and spectroscopic variability, it would relate with $f_{\mathrm{rot}}$ instead of $f_{\mathrm{env}}$.  Thus, the magnetosphere is unlikely to produce the found variability.

Periodic (increased) mass loss could also lead to periodic circumstellar variability.  Various processes could lead to increased mass loss, yet only a limited amount would do this in a periodic manner.  Beating effects between NRPs would be such a process, similar to the mass-loss mechanism responsible for the Be phenomenon \citep[e.g.,][]{1998cvsw.conf..207R, 2003A+A...411..229R, neinermathis2013, 2016A+A...588A..56B}.  However, we did not detect any stellar pulsations for \ZetOri\,Aa.

\subsection{Remaining variability}
\label{sec:discussion_instrumental}
In addition to variability related to the stellar rotation or the circumstellar environment, we also extracted several other frequencies. These do not seem to fall in a particular frequency domain with a straightforward interpretation, and show some sort of relation through (simple) combinations. We investigated these frequencies and chose $f_{\mathrm{x}} = 0.036 \pm 0.003$\,\d to be an independent frequency, because it explains the majority of the remaining frequencies.  However, since the origin of this periodic variability is not understood, a different frequency of the combinations might be the proper, independent one.  Nevertheless, to produce frequency combinations, the fundamental frequencies (and their combinations) need to originate from the same component, i.e., \ZetOri\,Aa.  Therefore, our assumption that all variations arise from \ZetOri\,Aa is likely correct.

Since \ZetOri\,Aa is undergoing a rather turbulent stage of its evolution, it is also possible that parts of the non-periodic variability are produced by the star itself.  A possible mechanism could then be related to the subsurface convective layer.  However, the highly variable H$\alpha$ line, used as one of the diagnostics for the circumstellar environment, does not seem to support this.  Additional (spectroscopic) measurements are needed to find observational support for this origin of the non-periodic photometric variability.

\section{Conclusions}
\label{sec:conclusions}
We performed a detailed time series analysis on the individual red and blue BRITE photometry of the magnetic, massive supergiant \ZetOri\,Aa, which we subdivided into three distinct epochs based on both observational evidence and data characteristics. In addition, we also performed a frequency analysis of the combined BRITE photometry, ignoring the color information of the observations. In total, we extracted 16 different frequencies, which were present in at least two of the six studied lightcurves. These 16 frequencies can be explained by three independent frequencies, their higher harmonics or combinations. For two of the independent frequencies, we understand their stellar origin, stellar rotation ($f_{\mathrm{rot}}$) and the cirumstellar environment ($f_{\mathrm{env}}$). A third independent frequency ($f_{\mathrm{x}}$) is necessary to explain the remaining variability as frequency combinations.

One family of frequencies corresponds to the rotation period of \ZetOri\,Aa, most likely related to two or four brightness spots, evenly distributed in longitude over the stellar surface. Therefore, we were able to determine the rotation period, $P_{\mathrm{rot}}=6.82\pm0.19$\,d, compatible with the literature value of $6.83\pm0.08$\,d \citep{2015A+A...582A.110B}, within the error bars. We cannot fully exclude that a large number of spots is present at the stellar surface of \ZetOri\,Aa, mimicking a few (equidistant) spots. However, the presence of a dipolar magnetic field often implies a simple spot geometry related to the magnetic poles.

The presence of the fourth harmonic of $f_{\mathrm{rot}}$ supports the recurrence timescale for DACs, $t_{\mathrm{rec}}=1.67\pm0.06$\,d, consistent with that of \citet{1999A+A...344..231K}, i.e., four CIRs are related to enhanced brightness regions evenly distributed in longitude. The mechanism that creates and sustains these inhomogeneities is still unknown, a beating effect between NRPs or a quadrupolar component to the stable dipolar magnetic field being the preferred pathways. If they are created by a stable magnetic field, a significant quadrupolar component, marginally compatible with the results of \citet{2015A+A...582A.110B}, is needed. Moreover, we did not detect any NRPs for \ZetOri\,Aa, making it impossible to corroborate this mechanism as the source for the four regions.

Thanks to the combination of the BRITE photometry and the simultaneous ground-based, high-resolution optical CHIRON spectroscopy of H$\alpha$ and \ion{He}{I} lines, we linked the $10.0\pm0.3$\,d variability to the circumstellar environment. This period is more pronounced during certain regions of the BRITE lightcurves dominated by the variation of the circumstellar environment. In addition, it explains the noted differences between the three observing epochs during the two BRITE monitoring campaigns. This periodic or cyclic variability of the circumstellar environment is unlikely to be related to the weak magnetosphere of \ZetOri\,Aa since the confined magnetosphere variability would instead be related to the stellar rotation. Periodically enhanced mass loss similar to Be stars, through beating of NRPs, might be an alternative driving mechanism for the observed variability. Finally, the strong variability of the circumstellar environment might be an explanation for the discrepancies between the 2011--12 Narval spectropolarimety and the dipolar fit of \citet{2015A+A...582A.110B}.

For the presence of enhanced brightness regions, leading to DAC variability, and for periodic mass loss, explaining the circumstellar variability, a beating between NRPs is proposed as a driving mechanism. However, we do not identify pulsations in the BRITE photometry of \ZetOri\,Aa.

A high-cadence detailed UV and visual spectropolarimetric campaign, simultaneously with high-precision, long-baseline photometry will be necessary to fully unravel the origin of the variability of \ZetOri\,Aa, and to determine the possible effects on its circumstellar environment and its magnetic field. Only then, will it be possible to investigate \ZetOri\,Aa for possible NRPs.

\begin{acknowledgements}
This research has made use of the SIMBAD database operated at CDS, Strasbourg (France), and of NASA's Astrophysics Data System (ADS). 
N.\,D.\,R. is grateful for postdoctoral support by the University of Toledo and by the Helen Luedke Brooks Endowed Professorship.
T.\,R. acknowledges support from the Canadian Space Agency grant FAST.
A.\,F.\,J.\,M. and G.\,A.\,W. acknowledge financial support from NSERC (Canada). A.\,F.\,J.\,M. also acknowledges support from FQRNT (Quebec).
R.\,E.\,M. acknowledges support from VRID-Enlace 214.016.002-1.0 and the BASAL Centro de Astrof{\'{i}}sica y Tecnolog{\'{i}}as Afines (CATA) PFB--06/2007.
G.\,H. wishes to acknowledge partial financial support from the Polish NCN grants 2015/18/A/ST9/00578.
A.\,P. acknowledges support from the Polish NCN grant 2016/21/D/ST9/00656.
R.\,K. and W.\,W.\,W. acknowledge support by the Austrian Research Promotion Agency (FFG)
Part of the research leading to these results has received funding from the European Research Council (ERC) under the European Union’s Horizon 2020 research and innovation programme (grant agreement N$^\circ$670519: MAMSIE).
The work was supported by the BRITE PMN grant 2011/01/M/ST9/05914.
B.\,B. wishes to thank A.\,Pigulski for the numerous comments and suggestions which greatly improved this work.

\end{acknowledgements}
\bibliographystyle{aa}
\bibliography{PhD_ADS}

\newpage
\begin{appendix}
\section{Data reduction of BRITE-photometry}
\label{sec:appendix_correction}
Since the data reduction and preparation of BRITE photometry is not trivial, we provide some additional information on its processing. The points raised here mainly originate from the treatment of \ZetOri itself, but can be generalized for most targets observed by the BRITE-Constellation. We decided to make the dedicated \textit{Python} routines publicly available to the community so they could be used and improved upon to serve for all different data releases. These routines are available for download on Github\footnote{\url{http://github.com/bbuysschaert/BRITE_decor}}, and adopt our previous experience with space photometry and the BRITE Cookbook by A.\,Pigulski\footnote{\url{http://brite.craq-astro.ca/doku.php?id=cookbook}} \citep[see also][]{2016A+A...588A..55P}. Several issues are treated by the developed routines. First, a timing issue occurs, most pronounced for certain observing modes during some of the earlier observing campaigns. Next, we expand upon the different types of data outliers encountered and how they are treated. Third, we discuss the various instrumental correlations we observed. Here, we put an additional focus on the temperature dependent point spread function (PSF) shape changes, causing a time dependent instrumental correlation between the observed signal and the CCD centroid positions. Finally, we give a few comments on the merging of different BRITE datasets.

\subsection{Mid-exposure times}
During some of the earlier observing campaigns by the BRITE-Constellation (including parts of the Orion\,II campaign), experiments were conducted on the feasibility and usefulness of onboard stacked data. As such, several exposures were merged on board, and once downlinked, they were considered as an individual datapoint. The released data, however, only provide the time at the beginning of the exposure, making it difficult to merge various datasets with distinct exposure times and a various amount of stacked images, i.e., with different durations. To this end, we determined the mid-exposure time for all observations, permitting an easier combination of datasets. These mid-exposure times were deduced for the onboard reference frame of the nano-satellite before applying the heliocentric correction.

In addition to the exposure time, the different time intervals between consecutive stacked images were accounted for. Combining this information leads to
\begin{equation}
\mathrm{JD_{mid}} = \mathrm{JD_{start}} + \frac{\mathrm{t_{exp}} \cdot N + \mathrm{t_{stack}} \cdot (N -1)}{2} \ \mathrm{,}
\label{eq:timing_correction}
\end{equation}
\noindent where $\mathrm{JD_{mid}}$ and $\mathrm{JD_{start}}$ are the mid-exposure and start of the exposure Julian Date, respectively. The exposure time is given by $\mathrm{t_{exp}}$, the time difference between consecutive stacked images by $\mathrm{t_{stack}}$, and $N$ is the number of stacked images corresponding to the observation. Typically, $\mathrm{t_{exp}}$ is 1\,s.  Different settings were accepted for the onboard stacking, but were kept constant for the duration of a given setup.  Typically, the adopted $\mathrm{t_{stack}}$ varies between 13\,s and 21\,s while $N$ ranges from 1 to 5.  Therefore, the applied correction can add up to 30\,s. Although the correction on the timing is minimal for non-stacked data, we consider it good practice to calculate and apply it nonetheless. We determined and applied this correction for all BRITE photometry, as a first step in the preparation process. 

\subsection{Outlier removal}
Here, we provide an overview on the different types on spurious datapoints we encountered during our analysis of the measurements of \ZetOri. In addition, several remarks are made for some of the outliers, and suggestions given for an effective treatment. This process was performed on the full datafile corresponding to one observational setup of a given BRITE nano-satellite. In the case of strongly gapped measurements, different parts can be treated separately.

The first type of spurious signal is caused by an incomplete rendering of the aperture on the saved CCD raster. Because of the movement of the target over the CCD raster, some of the signal was lost. These observational points were marked during the extraction since data release 2; therefore, remove these flagged datapoints for the remainder of the analysis.

\begin{figure}[t!]
		\centering
			\includegraphics[width=\columnwidth, height = 0.33\textheight]{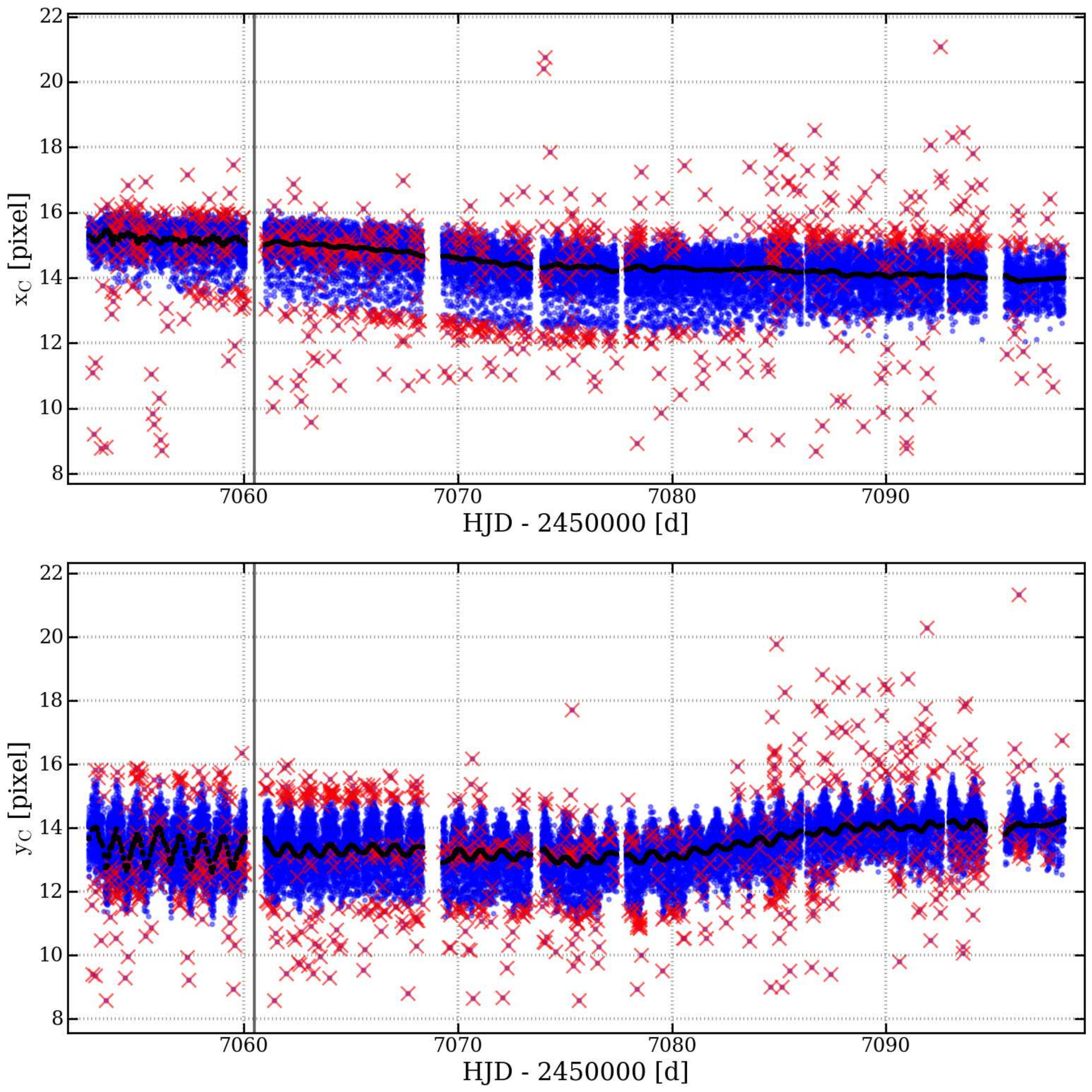}%
			\caption{Behavior of the CCD centroid positions with respect to time for the setup 6 file of BLb during the Orion II campaign at blue wavelengths. The model for the behavior is marked in black, while the merged outliers from both centroid positions are marked by the red crosses. These outliers occur in the outer $0.8\%$ of the model corrected centroid positions. This data file was corrected with different models for each one of the two distinct regions, indicated by the vertical line at HJD=2457060.5d. }
			\label{fig:appendix_position_outlier}
\end{figure}

Next, we studied the behavior of the CCD centroid position with respect to time. Not only does the centroid position have a clear relation with respect to time, but there are also outliers with respect to this trend. In the majority of cases, these outliers show anomalous behavior along both CCD dimensions and in flux, indicating that their removal is necessary. We accounted for the relation between the centroid position and time by means of a local linear regression fit.  The outliers themselves were identified by studying the distribution of the residual centroid positions.  We applied the rejection scheme for both CCD dimensions simultaneously, because often measurements were found to be an outlier along both dimensions (and in flux).  Additionally, this also ensured a homogeneous treatment of both CCD dimensions.  Lastly, we note that a rejection slightly stronger than that suggested by a visual inspection is favored. Typically, we aimed to reject up to the outer 5\,\% of the residual positions. This permitted an easier and clearer correction for any instrumental effects during subsequent steps, in particular for the PSF fluctuations. An example of the outlier rejection process on the CCD centroid positions is shown in Fig.\,\ref{fig:appendix_position_outlier} for the setup 6 file of BLb during the Orion II campaign.

\begin{figure}[t!]
		\centering
			\includegraphics[width=\columnwidth, height = 0.33\textheight]{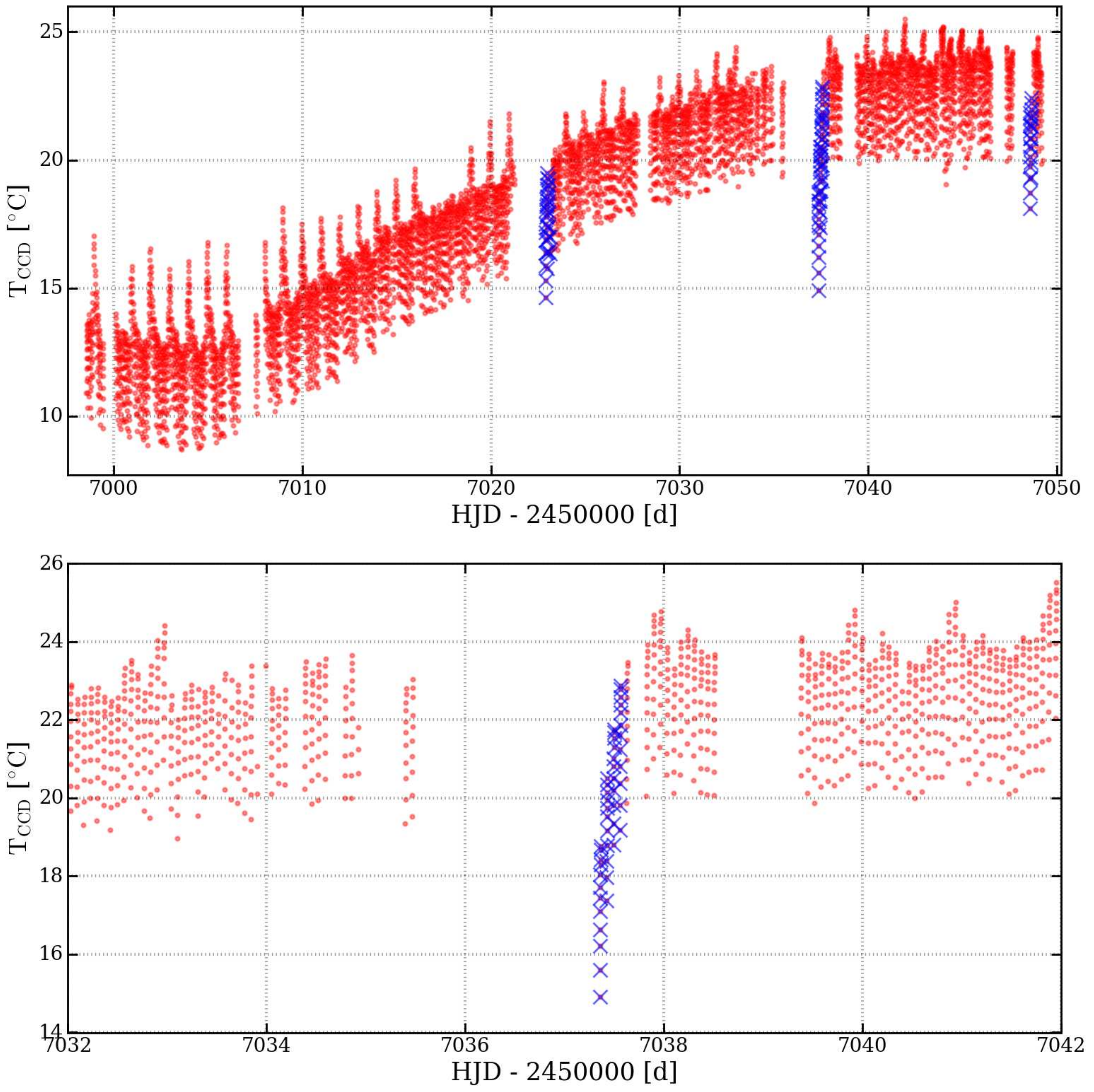}%
			\caption{\textit{Top}: Varying onboard temperature, measured at the CCD, is indicated in red for the setup 5 file of BHr during the Orion II campaign. The orbits having a different long-term behavior are rejected and indicated by blue crosses, which mostly happens after a significant data gap. \textit{Bottom}: Zoom of the top panel around the second marked region at HJD=2457037.}
			\label{fig:appendix_temperature_outlier}
\end{figure}

The subsequent type of spurious signal we removed is due to a differently behaving CCD temperature, $T_{\mathrm{CCD}}$, with respect to time. In addition, the temperature for satellite orbits just after significant time gaps is slightly different from the long-term relation of temperature with time. We show an example of this behavior in Fig.\,\ref{fig:appendix_temperature_outlier}. There are several satellite orbits with a lower temperature than expected (even when accounting for variations with a period of 1\,d). We aimed to be as conservative as possible, yet we removed these measurements to aid the subsequent instrumental correction process. Hence, we marked such datapoints for exclusions with a Graphical User Interface.

Finally, we flagged any remaining spurious flux measurements as outliers. The exact reason for such anomalous behavior can range from mild charge transfer inefficiency (CTI) to actual stellar signal \citep{2016PASP..128l5001P}. Since a basic or rough model is needed to represent the stellar signal (e.g., for pulsators or eclipsing binaries) before such outliers become apparent, this is an iterative process, that is heavily tweaked for the star under examination.

With all outliers removed, we performed one final test before applying instrumental corrections. Here, we investigated the number of observations per satellite orbit, and forced the threshold to correspond to a pre-set minimum. This permitted us to treat all satellite orbits in a coherent manner, and avoid influence from small number statistics. Typically, we aimed to have at least ten photometric measurements per orbit passage, while accounting for onboard stackings of consecutive images.

\subsection{Temperature-dependent PSF fluctuations}
\label{sec:appendix_PSFdetrend}
\begin{figure}[t!]
		\centering
			\includegraphics[width=\columnwidth, height = 0.33\textheight]{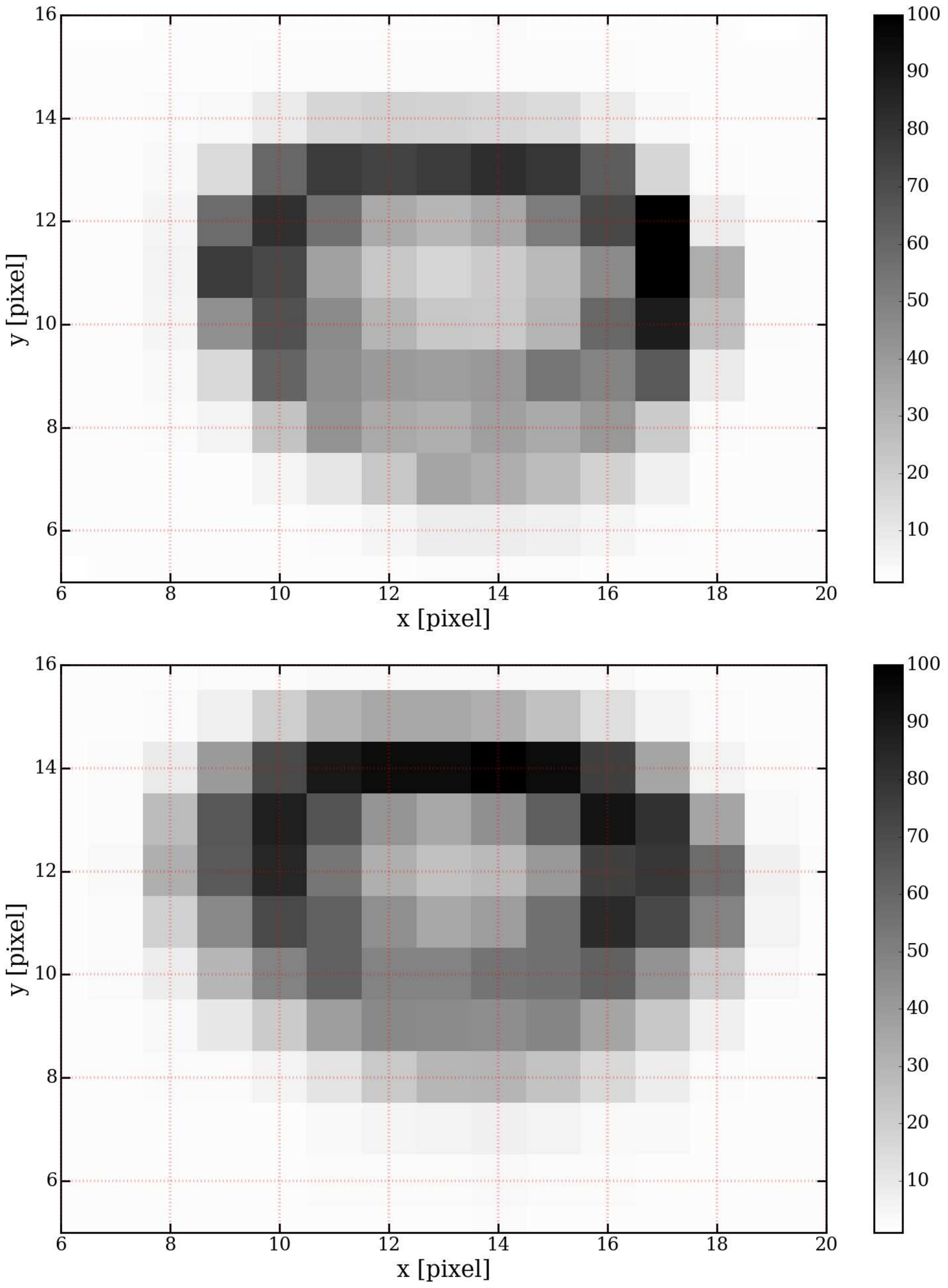}%
			\caption{Two different observations of \ZetOri during the setup 7 Orion\,II campaign for BHr. The top image was taken with an onboard temperature of 21.49\,$\mathrm{ ^{\circ}C}$, while the bottom image has a $T_{\rm CCD}$ of 17.57\,$\mathrm{ ^{\circ}C}$. The normalized counts per pixel is represented by the grayscale. Differences are seen in the width, the sharpness, and the location of the maximum of the non-Gaussian PSF.}
			\label{fig:appendix_PSF_changing}
\end{figure}

We started from the outlier-rejected photometry to correct for any possible instrumental trends or correlations. During our study of the BRITE photometry of \ZetOri, we noted one particular correlation that was much stricter than all the others and present in almost all datasets. It is now understood that this instrumental effect was caused by minor distortions in the lenses of the optical design, due to the fluctuating onboard temperature distorting the shape of the PSF.  This PSF has a shape which heavily depends on the orientation with the optical axis, and strongly differs from an airy disk for targets that fall close to the edge of the CCD \citep[see Fig.\,6 of][for the effect on the PSF with the orientation with respect to the optical axis]{2016PASP..128l5001P}.  Therefore, the time dependent temperature variations produce a time dependent PSF shape, causing a correlation between flux and the CCD centroid position of the PSF (which was the provided metadata diagnostic). This is most likely further enhanced by intra-pixel variations. In Fig.\,\ref{fig:appendix_PSF_changing}, we indicate two different CCD images for \ZetOri for the BHr setup 7 data, with different onboard temperatures (measured at the CCD). Since the centroid position is the average position of this shape on the CCD, one indeed expects the observed correlation between position and CCD temperature (and eventually flux). In the following, we discuss our correction procedure for this instrumental trend.

\begin{figure}[t!]
		\centering
			\includegraphics[width=\columnwidth, height = 0.17\textheight]{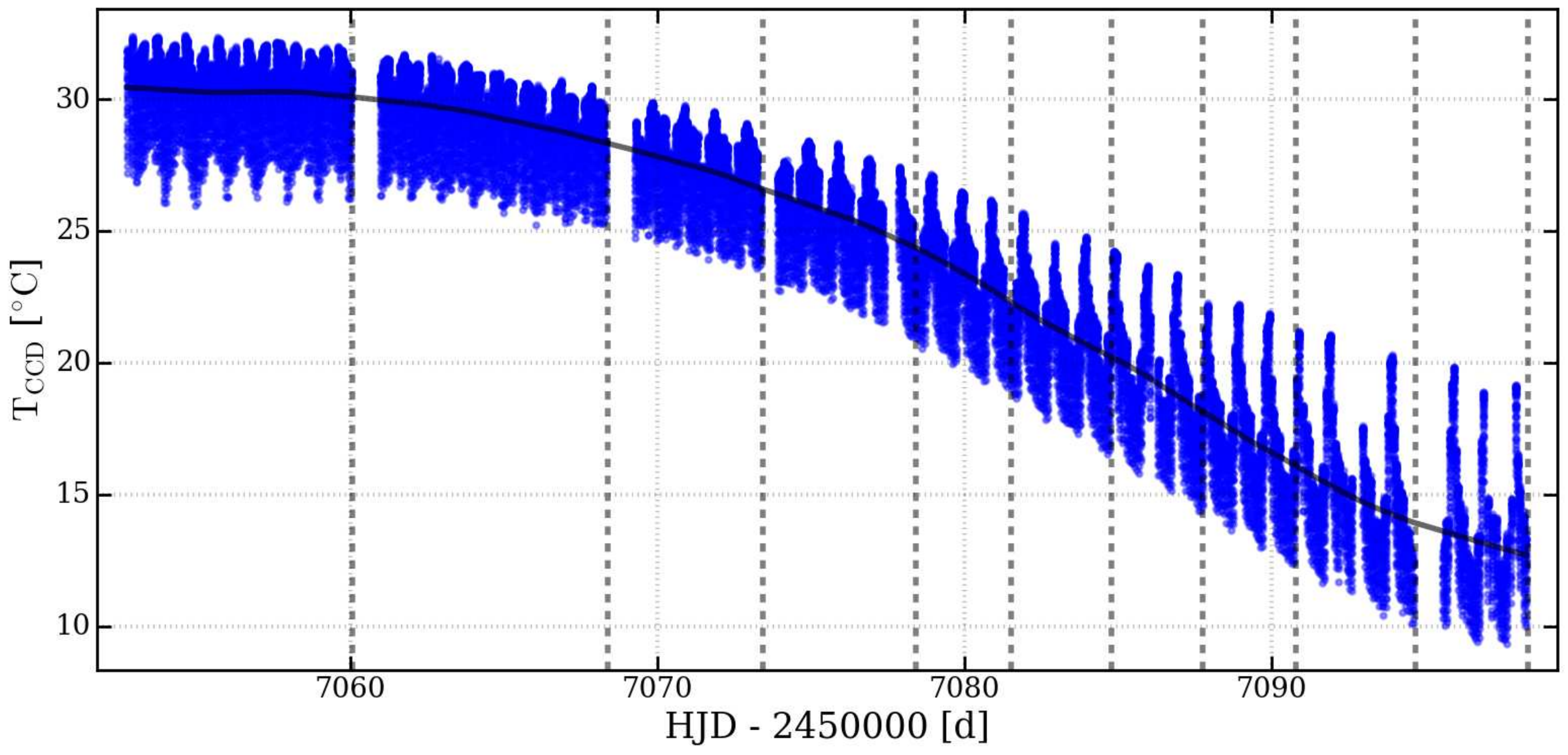}%
			\caption{Temperature behavior with respect to time for the setup 6 data of the Orion\,II campaign of BLb. The long-term behavior is approximated by a local linear regression filter, which constructs a total of ten different time windows using both a temperature threshold of 2$\degree$ and a datagap criterion of 0.5\,d. The correlation for the first window is shown in Fig.\,\ref{fig:appendix_PSFdetrend_effect}.}
			\label{fig:appendix_temperature_bins}
\end{figure}

As the temperature on board the BRITE nano-satellites is highly variable with time, the correlation between flux and centroid position is smeared out when the full data of a given observational setup are investigated. Therefore, it is best to discretize the dataset, so that the trend becomes more apparent within each individual subset. We constructed these subsets, or time windows, by studying the long-term temperature behavior with respect to time, indicated in Fig.\,\ref{fig:appendix_temperature_bins} for BLb data of \ZetOri. This long-term behavior was approximated by a local linear regression of the actual CCD temperature. From this trend, we defined a time window when the temperature difference was either larger than the set threshold, typically 1$\degree$ to 2$\degree$, or when the size of a data gap was too large. These limits were then slightly refined in a consecutive step in order not to treat the same satellite orbit passage differently. We include the limits of these windows in Fig.\,\ref{fig:appendix_temperature_bins}.

Within each time window, we intended to determine a description for the instrumental flux. Since the centroid positions are the only two parameters provided to study the PSF \citep[for most data releases; the more recent releases have two additional parameters, see also][]{2016A+A...588A..55P}, we use these as fitting variables. This leads to the approximation
\begin{equation}
\begin{split}
F(t) &= F_{\rm inst}(t) + F_{\rm star}(t) \\
     &\approx F_{\rm inst}(\mathrm{PSF}(T_{\rm CCD}(t))) + F_{\rm star}(t) \\
     &\approx F_{\rm inst}(x_{\rm c}(t),\,y_{\rm c}(t)) + F_{\rm star}(t) \ \mathrm{,}
\label{eq:PSF_flux_approximation}
\end{split}
\end{equation}
\noindent where $F(t)$ is the total flux, containing the instrumental flux, $F_{\rm inst}(t)$, and the stellar signal, $F_{\rm star}(t)$. We assume that we can approximate $F_{\rm inst}(x_{\rm c}(t),\,y_{\rm c}(t))$ as $F_{\rm inst}(x_{\rm c}(t)) + F_{\rm inst}(y_{\rm c}(t))$, which is only fully valid if both terms are uncorrelated. This simple approximation permits the usage of one-dimensional B-splines to recover $F_{\rm inst}(x_{\rm c}(t))$ and $F_{\rm inst}(y_{\rm c}(t))$ individually. The constant length of the fitting regions along the centroid position and the order of the spline are also considered as free parameters during the fitting process. We use both the Aikaike Information Criterion (AIC) and Bayesian Information Criterion (BIC)\footnote{Because of the different nature of the two criteria, they might favor a slightly different solution. When this happened, we used the best $\Lagr(\Theta)$ value to guide us between the proposed solutions.}, given as 
\begin{equation}
\mathrm{AIC}(\Theta) = 2 k - 2 \Lagr(\Theta) \ \mathrm{,}
\label{eq:AIC}
\end{equation}
\begin{equation}
\mathrm{BIC}(\Theta) = -2 \Lagr(\Theta) + k \ln(N) \ \mathrm{,}
\label{eq:BIC}
\end{equation}
\noindent to determine the optimal fit and the dominance of the respective CCD axes. Here, $N$ is the total number of datapoints within the given time window, $k$ is the number of estimated fitting parameters, given as 
\begin{equation}
k = (n+1) \cdot (o + 1) \ \mathrm{,}
\label{eq:estimated_parameters_IC}
\end{equation}
\noindent where $n + 1$ is the total number of defined fitting regions, and $o$ the order of the B-spline, respectively. We define the log-likelihood, $\Lagr(\Theta)$, as
\begin{equation}
\Lagr(\Theta) = \sum \limits^{}_{i} \left\{\ln \left(|\Magr(\Theta; i)|\right) + \frac{|\Dagr(i)|}{|\Magr(\Theta; i)|}\right\}
\label{eq:log_likelihood}
\end{equation}
\noindent for a parameter set $\Theta~=~\left(\Theta_1, \Theta_2, \dots, \Theta_k \right)$, dataset $\Dagr$, and model $\Magr(\Theta)$ \citep[see e.g.,][]{1986ssds.proc..105D, 1990ApJ...364..699A}. In our case, $\Dagr$ contains the flux measurements as a function of a centroid position, and $\Magr(\Theta)$ its B-spline representation, within a given time window.

\begin{figure}[t!]
		\centering
			\includegraphics[width=\columnwidth, height = 0.33\textheight]{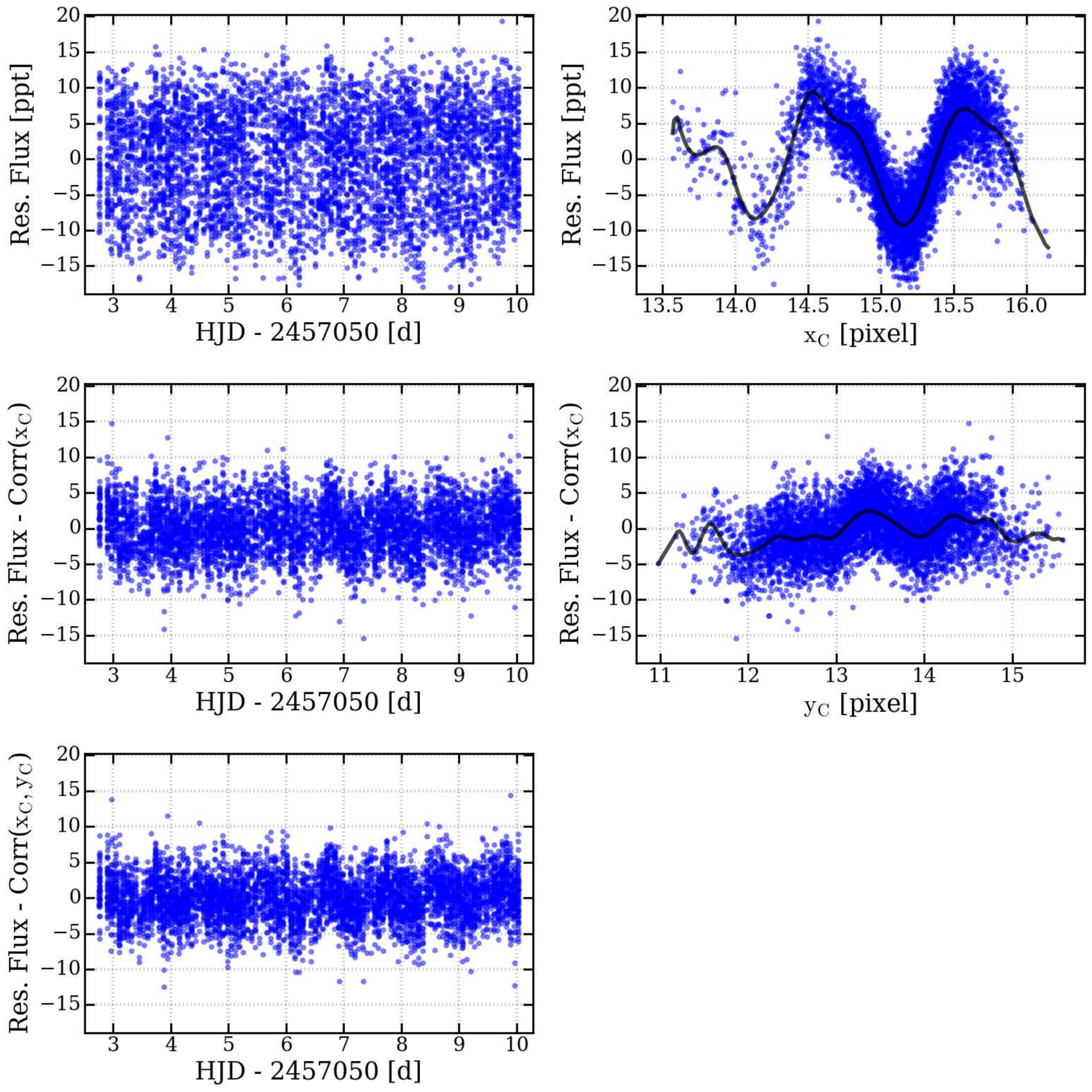}%
			\caption{\textit{Top left}: BLb residual flux of \ZetOri given in blue, from the first correction window of Fig.\,\ref{fig:appendix_temperature_bins}, which was subtracted by a local linear regression of the flux acting as a model for the stellar signal, $F_{\rm star}(t)$, and for the long-term instrumental signal. \textit{Top right}: Same residual flux as a function of the CCD centroid position $x_c$ indicating the correlation, which is approximated by the B-spline given in black. The correlation between flux and $x_c$ is more dominant than with $y_c$, hence it is decorrelated first, as indicated by the AIC and BIC. \textit{Middle left}: Same BLb flux residuals, now corrected for the correlation with $x_c$. \textit{Middle right}: Correlation between the corrected flux residuals and $y_c$, together with the B-spline representation in black. \textit{Bottom left}: Final flux residuals, corrected for the correlation with both CCD centroid positions, ready to correct for remaining instrumental effects. The periodic variations in this final product is caused by the periodic $T_{\rm CCD}$ fluctuations (see Fig.\,\ref{fig:appendix_temperature_bins}).}
			\label{fig:appendix_PSFdetrend_effect}
\end{figure}

Typically, we found the B-spline to be of order 3, and the constant knotpoint spacing between 0.1 and 0.5\,pixel. An example for such a spline representation and its effect on the BRITE-photometry is given in Fig.\,\ref{fig:appendix_PSFdetrend_effect} for the first time window of the BLb data of \ZetOri for setup file 6 shown in Fig.\,\ref{fig:appendix_temperature_bins}. Again, we recommend subtracting an approximative model for the stellar flux, $F_{\rm star}(t)$, while performing the decorrelation process in an iterative manner.

Figure\,\ref{fig:appendix_PSFdetrend_effect} indicates that the main effect of this correction is a significant reduction of the instrumental noise. This noise reduction ranges from roughly a factor of 1.2 in the case of onboard stacked photometry and up to a factor of 2.5 for the non-stacked data. Thus, we encourage the application of this correction before remedying other instrumental effects.

\subsection{Instrumental signal decorrelation}
The final step in the preparation of the BRITE observations accounted for any remaining instrumental effects. Here, we consider four different possible observables that can correlate with the measured signal: i) the onboard temperature, $T_{\rm CCD}$, measured at the CCD; ii) the x-dimension and iii) the y-dimension of the CCD centroid position; and iv) the satellite orbital phase, $\phi_s$. The most recent data releases of BRITE photometry provide two additional observables, indicating the sharpness of the PSF and the variations of the background of the CCD (Popowicz et al., in prep.).  These can also be employed for this decorrelation process. Currently, the full data per observational setting of a given satellite is decorrelated as a whole, but, except for $T_{\rm CCD}$, the decorrelation process can be performed on smaller subsets of the data. An example of this was discussed in the previous section. In addition, working on data subsets can also aid the decorrelation with $\phi_s$ in certain special cases. However, we advise against doing so for $T_{\rm CCD}$, since it complicates the correction for the long-term temperature trend with respect to time.

To determine the most suitable correcting sequence, we use the normalized correlation matrix for $\left[\mathrm{F}, T_{\mathrm{CCD}}, x_c, y_c, \phi_s\right]$. We correct for the dominant correlation between flux and a diagnostic parameter first, but only when it is greater than $10\%$ in absolute value. If the correlation does not pass the set threshold, the process ends. Otherwise, we employ a similar B-spline fitting scheme as in Sect.\,\ref{sec:appendix_PSFdetrend} to represent and correct for the correlations. Once corrected, we recalculate the correlation matrix, and search for the next most significant correlation to repeat the procedure.

For the decorrelation process, we mention several caveats. First, we permit each diagnostic parameter to be used only once, if at all. Repetitive fitting with and correction for the same parameter leads to overfitting. Second, if the CCD temperature was not used for the correction, we enforced this for the case of \ZetOri since we were mainly interested in the very low-frequency regime ($ < 0.5$\,\d). Hence, any possible effect of the long-term temperature behavior should be accounted for. Next, because the adopted information criteria penalize representations with too many free parameters, we tried to avoid over-describing the data. Nevertheless, sensible limits should be provided to the B-spline fitting algorithm. Finally, the correlation matrix is only sensitive to linear correlations between the different parameters. Therefore, we suggest a final, visual test to ensure no remaining correlation is present in the final data product.

For our processing of \ZetOri, we only needed to correct for the onboard temperature, or we needed to enforce it. Only on a few occasions did we deduce a significant correlation with the satellite orbital phase. Thanks to the decorrelation process for the varying PSF shape, no significant correlation remained between the observed flux and the centroid positions.

Finally, after completion of the correction process, we performed a final outlier rejection for the corrected flux. As such, possible significantly altered flux is dealt with in a coherent manner. These flux outliers were flagged with a sigma filter.

\subsection{Merging different datasets}
Merging of different BRITE datasets, either from different nano-satellites or from the same one, is most often straightforward when the same exposure time and number of onboard stacking, $N$, were used. In the case of differences between the onboard stacking, the data should be treated with care, and detailed techniques or weighting are advised. A common method is to merge each satellite passage to one individual datapoint \citep[e.g.,][]{2016A+A...588A..56B}. This is only advisable, however, when one is not interested in the high-frequency region (i.e., above $f_{\mathrm{sat}}/2$), the low-frequency regime (i.e., below $\sim$0.5\,\d) or when only considering model fitting. Otherwise, proper weighting is mandatory, accounting for differences in $N$ and preferably differences in data quality \citep[e.g.,][]{2003BaltA..12..253H}. There is no unique method for combining these techniques, hence the weighting method will most likely be decided on a star-by-star basis.

\section{Tables}
Here, we provide some more information related to the extracted frequencies, presented in Table\,\ref{tab:BRITE_frequencies}. Table\,\ref{tab:BRITE_amplitudes} overviews the amplitudes of the individual extracted frequencies. The confidence interval on the determined amplitudes is calculated using
\begin{equation}
\delta A = \sqrt{\frac{2}{N}} \delta N \ \mathrm{,}
\label{eq:uncertainty_amplitude}
\end{equation}
for which $N$ is the number of measurements, and $\delta N$ is the standard deviation on the final flux residuals \citep[][]{1999DSSN...13...28M}.  The considerable number of corrections applied to the BRITE photometry also has an impact on the noise characteristics of the data, causing this formalism to be a strong underestimation of the real uncertainty.  The S/N of the individual frequencies are shown in Table\,\ref{tab:BRITE_SNR}.

Information about the different CHIRON observations is given in Table\,\ref{tab:CHIRON_log}.

\begin{table*}[h!t]
\caption{Amplitudes of the extracted frequencies describing the photometric variability of \ZetOri seen in the studied BRITE lightcurves.  The first column of each dataset presents the amplitude of the extracted frequencies for the full dataset, while the second column is for the non-periodic-event excluded photometry. Uncertainties on the amplitudes were calculated using Eq.\,(\ref{eq:uncertainty_amplitude}).  All amplitudes are given in parts per thousand.}
\centering
\tabcolsep=1pt
\begin{tabular}{p{2.5cm}||c||cccc||cccc||cccc}
\hline
\hline
& Combined & \multicolumn{4}{c||}{Orion\,I} & \multicolumn{4}{c||}{Orion\,IIa} & \multicolumn{4}{c}{Orion\,IIb}\\
Frequency &  & \multicolumn{2}{c}{Blue} & \multicolumn{2}{c||}{Red} & \multicolumn{2}{c}{Blue} & \multicolumn{2}{c||}{Red} & \multicolumn{2}{c}{Blue} & \multicolumn{2}{c}{Red}\\
$\delta A$	&$\pm	0.14 	$&$\pm	0.06	$&$\pm	0.07	$&$\pm	0.05	$&$\pm	0.06	$&$\pm	0.10	$&$\pm	0.13	$&$\pm	0.10	$&$\pm	0.11	$&$\pm	0.05	$&$\pm	0.07	$&$\pm	0.06	$&$\pm	0.07	$\\
\hline
$f_{\mathrm{rot}}$										&	2.14	&	1.92	&		&		&		&		&	2.81	&		&		&	3.13	&	2.66	&	4.13	&		\\
$2f_{\mathrm{rot}}$										&	3.07	&	4.26	&	3.82	&	4.50	&	4.63	&	2.13	&		&	2.97	&	3.77	&	2.05	&		&	5.19	&	3.22	\\
$3f_{\mathrm{rot}}$										&		&		&		&		&		&		&		&	2.61	&	3.02	&	2.44	&		&	3.04	&		\\
$4f_{\mathrm{rot}}$										&		&	2.53	&		&	1.86	&		&	2.85	&	3.48	&	3.12	&	3.35	&		&	2.08	&		&		\\
&&&&&&&&&&&&&\\																											
$f_{\mathrm{env}}$										&	3.22	&	4.68	&	4.24	&	4.17	&	4.07	&	4.83	&		&	5.05	&		&		&		&	5.89	&		\\
$2f_{\mathrm{env}}$										&		&		&		&		&	2.46	&	4.32	&		&	4.95	&		&	2.46	&	2.29	&		&		\\
$4f_{\mathrm{env}}$										&		&		&		&	1.96	&		&	2.02	&	3.40	&		&		&	2.56	&		&		&		\\
&&&&&&&&&&&&&\\																										
$f_{\mathrm{x}}$											&	2.26	&	3.34	&		&	3.30	&		&		&	2.68	&	4.19	&	2.19	&		&		&	4.49	&		\\
&&&&&&&&&&&&&\\																										
$f_{\mathrm{rot}} + f_{\mathrm{env}}$					&	2.03	&	1.94	&		&		&		&	3.82	&	2.36	&	4.35	&		&		&		&		&		\\
$3f_{\mathrm{rot}} + f_{\mathrm{env}}$					&		&		&		&		&		&	2.53	&	2.10	&	2.80	&		&	2.48	&		&		&		\\
$4f_{\mathrm{rot}} + f_{\mathrm{env}}$					&		&		&		&		&		&		&		&		&	2.45	&		&		&	2.67	&	2.71	\\
&&&&&&&&&&&&&\\	
$f_{\mathrm{env}} + f_{\mathrm{x}}$						&	2.35	&	4.29	&	2.94	&	5.17	&		&	2.15	&		&		&		&		&		&		&		\\
$f_{\mathrm{env}} + f_{\mathrm{x}}$						&	3.24	&		&		&		&		&		&		&		&		&		&		&		&		\\
$f_{\mathrm{env}} + f_{\mathrm{x}}$						&	2.65	&		&		&		&		&		&		&		&		&		&		&		&		\\
$f_{\mathrm{env}} - f_{\mathrm{x}}$						&	2.35	&		&		&		&		&		&		&	2.99	&	4.41	&	4.53	&	3.55	&	3.47	&	4.40	\\
&&&&&&&&&&&&&\\	
$f_{\mathrm{rot}} - f_{\mathrm{x}}$						&	1.84	&		&		&		&		&		&		&		&		&		&		&		&		\\									$2f_{\mathrm{rot}} + f_{\mathrm{x}}$						&	1.88	&	1.84	&		&	3.08	&		&		&		&		&		&		&		&		&		\\
$2f_{\mathrm{rot}} - f_{\mathrm{x}}$						&	1.82	&		&		&		&		&		&		&	2.34	&	2.67	&	2.20	&		&	5.05	&		\\
\hline
\end{tabular}
\label{tab:BRITE_amplitudes}
\tablefoot{Most of the extracted frequencies are understood to be produced by three frequencies: $f_{\mathrm{rot}}$, corresponding to the stellar rotation, $f_{\mathrm{env}}$, related to the circumstellar environment; and $f_{x}$, a not-yet-understood frequency.}
\end{table*}

\begin{table*}[t]
\caption{S/N of the extracted frequencies describing the photometric variability of \ZetOri seen in the studied BRITE lightcurves.}
\centering 
\tabcolsep=3pt
\begin{tabular}{p{2.5cm}||c||cccc||cccc||cccc}
\hline
\hline
& Combined & \multicolumn{4}{c||}{Orion\,I} & \multicolumn{4}{c||}{Orion\,IIa} & \multicolumn{4}{c}{Orion\,IIb}\\
Frequency &  & \multicolumn{2}{c}{Blue} & \multicolumn{2}{c||}{Red} & \multicolumn{2}{c}{Blue} & \multicolumn{2}{c||}{Red} & \multicolumn{2}{c}{Blue} & \multicolumn{2}{c}{Red}\\
\hline
$f_{\mathrm{rot}}$										&	4.6	&	4.7	&		&		&		&		&	4.4	&		&		&	6.0	&	5.2	&	5.1	&		\\
$2f_{\mathrm{rot}}$										&	5.8	&	7.3	&	5.9	&	7.1	&	7.0	&	4.8	&		&	5.4	&	4.8	&	4.6	&		&	5.2	&	4.3	\\
$3f_{\mathrm{rot}}$										&		&		&		&		&		&		&		&	5.8	&	5.4	&	4.7	&		&	4.5	&		\\
$4f_{\mathrm{rot}}$										&		&	4.3	&		&	4.5	&		&	5.8	&	4.8	&	5.4	&	5.0	&		&	4.1	&		&		\\				
&&&&&&&&&&&&&\\																							
$f_{\mathrm{env}}$										&	6.0	&	7.6	&	7.3	&	6.9	&	6.7	&	8.6	&		&	6.4	&		&		&		&	6.4	&		\\
$2f_{\mathrm{env}}$										&		&		&		&		&	4.7	&	6.7	&		&	6.6	&		&	4.9	&	4.2	&		&		\\
$4f_{\mathrm{env}}$										&		&		&		&	4.3	&		&	4.5	&	4.7	&		&		&	4.8	&		&		&		\\
&&&&&&&&&&&&&\\																										
$f_{\mathrm{x}}$											&	5.4	&	5.3	&		&	6.3	&		&		&	4.4	&	5.5	&	4.4	&		&		&	4.8	&		\\
&&&&&&&&&&&&&\\																										
$f_{\mathrm{rot}} + f_{\mathrm{env}}$					&	4.8	&	4.3	&		&		&		&	5.3	&	4.5	&	6.7	&		&		&		&		&		\\
$3f_{\mathrm{rot}} + f_{\mathrm{env}}$					&		&		&		&		&		&	5.6	&	4.3	&	5.8	&		&	4.4	&		&		&		\\
$4f_{\mathrm{rot}} + f_{\mathrm{env}}$					&		&		&		&		&		&		&		&		&	4.4	&		&		&	4.1	&	4.2	\\
&&&&&&&&&&&&&\\	
$f_{\mathrm{env}} + f_{\mathrm{x}}$						&	5.3	&	4.1	&	5.2	&	7.2	&		&	4.1	&		&		&		&		&		&		&		\\
$f_{\mathrm{env}} + f_{\mathrm{x}}$						&	6.2	&		&		&		&		&		&		&		&		&		&		&		&		\\
$f_{\mathrm{env}} + f_{\mathrm{x}}$						&	4.6	&		&		&		&		&		&		&		&		&		&		&		&		\\
$f_{\mathrm{env}} - f_{\mathrm{x}}$						&	5.2	&		&		&		&		&		&		&	5.7	&	6.3	&	7.5	&	6.4	&	4.3	&	4.6	\\
&&&&&&&&&&&&&\\																										
$f_{\mathrm{rot}} - f_{\mathrm{x}}$						&	4.4	&		&		&		&		&		&		&		&		&		&		&		&		\\
$2f_{\mathrm{rot}} + f_{\mathrm{x}}$						&	4.6	&	4.2	&		&	5.6	&		&		&		&		&		&		&		&		&		\\
$2f_{\mathrm{rot}} - f_{\mathrm{x}}$						&	4.3	&		&		&		&		&		&		&	4.2	&	4.3	&	5.1	&		&	5.3	&		\\

\hline
\end{tabular}
\label{tab:BRITE_SNR}
\tablefoot{Most of the extracted frequencies are understood to be produced by three frequencies: $f_{\mathrm{rot}}$, corresponding to the stellar rotation, $f_{\mathrm{env}}$, related to the circumstellar environment; and $f_{x}$, a not-yet-understood frequency.}
\end{table*}

\begin{table*}[t]
\caption{Journal of the CTIO 1.5\,m/CHIRON observations of \ZetOri\,A.  Each spectrum consists of two co-added 8\,s observations.  The indicated S/N is measured at the continuum around the H$\alpha$ line.}
\centering
\tabcolsep=3pt
\begin{tabular}{cccccc|cccccc}
\hline
\hline
Nr. &	Date			&	UT		&	Mid-HJD 		& Seeing	& S/N	&	Nr. &	Date			&	UT		&	Mid-HJD 		& Seeing	& S/N	\\
	&				&			&	-2450000		&[arcsec]&		&		&				&			&	-2450000		&[arcsec]&		\\
\hline
1	&	2015-02-07	&	01:15:18	&	7060.555345	&	0.56	&	81	&	31	&	2015-02-25	&	01:51:19	&	7078.578924	&	0.92	&	88	\\
2	&	2015-02-07	&	03:25:05	&	7060.645468	&	0.86	&	94	&	32	&	2015-02-25	&	02:53:19	&	7078.621971	&	0.92	&	94	\\
3	&	2015-02-08	&	00:39:35	&	7061.530469	&	0.68	&	93	&	33	&	2015-02-26	&	02:05:12	&	7079.588471	&	0.92	&	99	\\
4	&	2015-02-08	&	03:17:17	&	7061.639974	&	0.90	&	98	&	34	&	2015-02-26	&	02:34:35	&	7079.608877	&	0.92	&	95	\\
5	&	2015-02-09	&	01:14:17	&	7062.554493	&	0.63	&	95	&	35	&	2015-02-27	&	01:56:58	&	7080.582671	&	1.11	&	94	\\
6	&	2015-02-09	&	03:17:46	&	7062.640235	&	0.54	&	100	&	36	&	2015-02-27	&	02:50:33	&	7080.619881	&	1.05	&	86	\\
7	&	2015-02-11	&	01:55:07	&	7064.582692	&	0.88	&	88	&	37	&	2015-03-04	&	00:29:06	&	7085.521220	&	0.90	&	96	\\
8	&	2015-02-11	&	04:24:01	&	7064.686092	&	0.60	&	73	&	38	&	2015-03-05	&	01:43:53	&	7086.573066	&	0.55	&	86	\\
9	&	2015-02-12	&	01:51:37	&	7065.580189	&	0.62	&	87	&	39	&	2015-03-05	&	02:26:10	&	7086.602417	&	0.47	&	82	\\
10	&	2015-02-12	&	04:10:03	&	7065.676312	&	0.93	&	78	&	40	&	2015-03-06	&	01:19:35	&	7087.556104	&	0.69	&	89	\\
11	&	2015-02-13	&	02:00:05	&	7066.585985	&	0.97	&	82	&	41	&	2015-03-06	&	02:55:20	&	7087.622590	&	0.55	&	82	\\
12	&	2015-02-13	&	02:52:05	&	7066.622102	&	0.94	&	86	&	42	&	2015-03-07	&	01:16:27	&	7088.553834	&	0.53	&	78	\\
13	&	2015-02-14	&	02:40:06	&	7067.613698	&	1.08	&	89	&	43	&	2015-03-07	&	02:28:51	&	7088.604116	&	0.63	&	77	\\
14	&	2015-02-14	&	03:27:30	&	7067.646611	&	0.94	&	73	&	44	&	2015-03-08	&	00:52:48	&	7089.537330	&	0.66	&	95	\\
15	&	2015-02-15	&	01:50:24	&	7068.579105	&	0.83	&	83	&	45	&	2015-03-08	&	02:11:50	&	7089.592212	&	0.66	&	85	\\
16	&	2015-02-15	&	03:29:21	&	7068.647824	&	0.55	&	86	&	46	&	2015-03-09	&	01:15:40	&	7090.553118	&	0.66	&	88	\\
17	&	2015-02-17	&	01:02:54	&	7070.545964	&	0.97	&	79	&	47	&	2015-03-09	&	02:29:04	&	7090.604082	&	0.66	&	81	\\
18	&	2015-02-17	&	01:52:09	&	7070.580162	&	0.61	&	90	&	48	&	2015-03-10	&	01:54:35	&	7091.580048	&	0.59	&	89	\\
19	&	2015-02-18	&	01:01:54	&	7071.545180	&	0.87	&	87	&	49	&	2015-03-10	&	02:29:23	&	7091.604207	&	0.85	&	80	\\
20	&	2015-02-18	&	01:24:50	&	7071.561120	&	0.75	&	96	&	50	&	2015-03-11	&	01:22:10	&	7092.557446	&	0.82	&	86	\\
21	&	2015-02-19	&	02:06:47	&	7072.590148	&	0.57	&	90	&	51	&	2015-03-11	&	02:38:55	&	7092.610737	&	1.29	&	93	\\
22	&	2015-02-19	&	03:13:17	&	7072.636337	&	0.59	&	83	&	52	&	2015-03-12	&	00:48:08	&	7093.533717	&	0.79	&	90	\\
23	&	2015-02-20	&	00:48:56	&	7073.536020	&	0.94	&	89	&	53	&	2015-03-12	&	02:07:38	&	7093.588928	&	0.57	&	80	\\
24	&	2015-02-21	&	01:21:11	&	7074.558330	&	1.06	&	83	&	54	&	2015-03-13	&	00:51:32	&	7094.535998	&	0.71	&	83	\\
25	&	2015-02-22	&	01:12:48	&	7075.552431	&	1.37	&	93	&	55	&	2015-03-13	&	02:31:15	&	7094.605229	&	0.77	&	88	\\
26	&	2015-02-22	&	03:31:39	&	7075.648848	&	1.41	&	78	&	56	&	2015-03-14	&	00:46:33	&	7095.532448	&	0.99	&	91	\\
27	&	2015-02-23	&	01:29:23	&	7076.563859	&	0.92	&	90	&	57	&	2015-03-15	&	00:45:08	&	7096.531369	&	0.93	&	87	\\
28	&	2015-02-23	&	03:16:12	&	7076.638031	&	0.83	&	88	&	58	&	2015-03-15	&	02:00:03	&	7096.583391	&	0.70	&	91	\\
29	&	2015-02-24	&	00:53:07	&	7077.538599	&	0.79	&	88	&	59	&	2015-03-16	&	00:51:42	&	7097.535847	&	0.41	&	95	\\
30	&	2015-02-24	&	03:05:38	&	7077.630605	&	0.84	&	81	&	60	&	2015-03-16	&	01:44:33	&	7097.572541	&	0.41	&	77	\\
\hline
\end{tabular}
\label{tab:CHIRON_log}
\end{table*}

\end{appendix}
\end{document}